\documentclass[twocolappendix,apj]{emulateapj}

\usepackage{perpage}
\usepackage{hyperref}

\shorttitle{Interaction between Cassiopeia A and Nearby Molecular Clouds}
\shortauthors{Kilpatrick, Bieging, \& Rieke}

\begin{document}


\title{Interaction between Cassiopeia A and Nearby Molecular Clouds}

\author{C. D. Kilpatrick\altaffilmark{1}, J. H. Bieging\altaffilmark{1}, \& G. H. Rieke\altaffilmark{1}}
\affil{Steward Observatory, University of Arizona, Tucson, AZ 85721}

\altaffiltext{1}{Steward Observatory, University of Arizona, 933 N. Cherry Ave, Tucson, AZ 85721 USA}


\begin{abstract}

We present spectroscopy of the supernova remnant Cassiopeia A (Cas A) observed at infrared wavelengths from $10-40$ $\mu$m with the \textit{Spitzer Space Telescope} and at millimeter wavelengths in $^{12}$CO and $^{13}$CO J $=2-1$ ($230$ and $220$ GHz) with the Heinrich Hertz Submillimeter Telescope.  The IR spectra demonstrate high-velocity features toward a molecular cloud coincident with a region of bright radio continuum emission along the northern shock front of Cas A.  The millimeter observations indicate that CO emission is broadened by a factor of two in some clouds toward Cas A, particularly to the south and west. We believe that these features trace interactions between the Cas A shock front and nearby molecular clouds.  In addition, some of the molecular clouds that exhibit broadening in CO lie $1 - 2'$ away from the furthest extent of the supernova remnant shock front.  We propose that this material may be accelerated by ejecta with velocity significantly larger than the observed free-expansion velocity of the Cas A shock front.  These observations may trace cloud interactions with fast-moving outflows such as the bipolar outflow along the southwest to northeast axis of the Cas A supernova remnant, as well as fast-moving knots seen emerging in other directions.

\end{abstract}


\keywords{infrared: ISM --- ISM: individual (Cassiopeia A) --- ISM: molecules --- ISM: supernova remnants --- shock waves}


\section{Introduction}

Supernova explosions are traditionally described in terms of a spherical shock that expands into the homogeneous interstellar medium.  This model ignores the kinematics of the actual supernova expansion where the boundary between the shock front and the interstellar medium is composed of material at varying temperatures and densities.  An important consequence of this more complex model of supernova physics is the evolution of shock fronts that encounter an interstellar cloud of size comparable to the supernova remnant (SNR).  Internal flows of the SNR are expected to become highly turbulent where the shock front interacts with a cloud \citep{jj99}.  Any spectral component should be velocity-broadened as a result of increased turbulence from a shock front/cloud interaction.  Magneto-hydrodynamic models indicate that the reverse shock behind the surface of interaction between the two media should be a site for particle acceleration and magnetic field amplification \citep{jj99}.  Consequently, a strong synchrotron component may be present between supernovae forward and reverse shocks.

The SNR Cassiopeia A (Cas A) is a nearby (3.4 kpc) and young ($\sim 350$ yr) example of a Type IIb supernova \citep{krause08} well-suited to observing these effects.  The proximity of the SNR and its inherent brightness make it ideal for studying the details of supernova physics including nucleosynthesis, light echoes, dust physics, and cosmic rays.  There is an extensive history of molecular spectroscopy of gas in the general direction of Cas A as well as the study of interactions between its ejecta and surrounding material \citep[e.g.,][]{weinreb63,barrett64,bc86,hs86,kra96,krause04}.  

Among this work, observations of carbon monoxide lines indicate that some of the gas toward the SNR is relatively warm ($T_{\mbox{\tiny K}} \sim 20$ K) \citep{wilson93}. However, whether there is an on-going interaction between the Cas A shock front and nearby molecular gas is ambiguous. Although CO line widths to the east and west may demonstrate some broadening \citep{ll99}, the 1720 MHz OH maser emission that is often seen in shocks is missing \citep{fm98}.  Perhaps the broadened CO results from an interaction that occurred recently and has not yet had time to develop more complex structure.  Given the youth of the SNR, the effects of interactions around Cas A in terms of warm molecular gas and broadened lines may be subtle.

Nonetheless, there are some enticing hints at an interaction.  The relatively warm temperature of the gas surrounding the SNR suggests an interaction between the shock front and molecular cloud along the western edge of Cas A.  In addition, \citet{arendt14} find that the dust at the outer edge of the remnant has a composition consistent with that of interstellar dust. \citet{fes11} identify variable knots of emission at a similar distance from the SNR center that they ascribe to interactions with ambient material.  Correlation between radio synchrotron emission and X-ray emission, especially toward the western edge of the remnant, suggest that the ejecta are moving into a denser medium while steep radio indices suggest relativistic particle acceleration \citep{ar96,kra96}.  These phenomena are consistent with an interaction between the fast-moving ejecta and a dense molecular cloud.  

\begin{figure*}
    \figurenum{1}
	\centering
		\includegraphics[width=0.44\textwidth,angle=270]{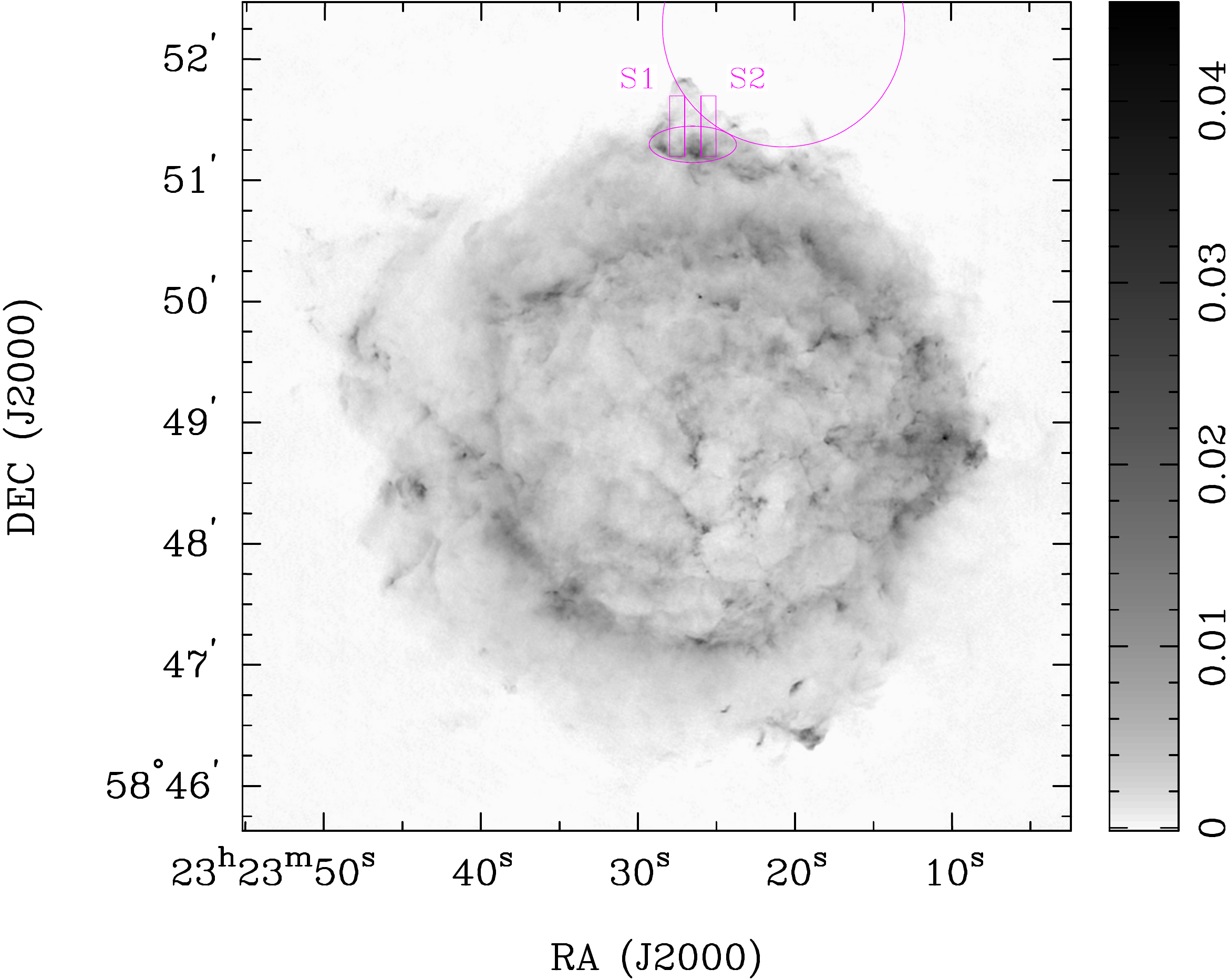}
	\caption{\scriptsize VLA 20 cm map of Cas A \citep{arlpb91} with positions of scan 1 (east) and scan 2 (west) overlaid as rectangular boxes (pink).  Each of these boxes represents 16 pointings of the \textit{Spitzer} IRS, with the southernmost pointing in each scan (i.e. S1, P1 and S2, P1) corresponding to the first pointing in each scan.  Each successive pointing in each scan is spaced $2''$ to the north of the previous pointing (i.e. S1, P2 is $2''$ due north of S1, P1).  The large circle (pink) represents the position of a molecular cloud identified in previous literature \citep{hines04}.  The ellipse at the south end of the two shock tracks outlines the region of bright radio continuum emission that suggests a shock front/cloud interaction.}
\end{figure*}

There is also a molecular cloud at the north shock front, with bright radio emission between this cloud and the Cas A reverse shock \citep[Figure 1;][]{hines04}.  The absence of any X-ray silicon emission or optical line emission at this position indicates that the radio emission does not result from an enhancement in the molecular gas density or from any obvious characteristic of the SNR.  Therefore, we infer that this radio emission may be a tracer for particle acceleration behind the point of interaction of the shock front and a molecular cloud \citep{jj99}.  Near-infrared K$_{s}$ emission has also been demonstrated to be dominated by a synchrotron continuum component \citep{rho03}.  The K$_{s}$ emission feature along the northern shock front \citep{rho03}, roughly coinciding with the enhanced radio emission, supports the possibility that we are seeing a SNR/MC interface. Enhancements in $K_{s}$ emission to the west and south indicate some other regions of possible interaction.

To search systematically for such interactions, we present radio and infrared spectroscopy of gas within several arcminutes of Cas A.  In particular, we looked for line broadening in the J $=2-1$ transitions of $^{12}$CO and $^{13}$CO.  We also searched for signs of an interaction in spectra taken from the \textit{Spitzer} Infrared Spectrograph (IRS) along two tracks in declination (Figure 1) across the northern shock front of Cas A.  These observations were performed using the high-resolution module aboard IRS, and given the integration time, are the deepest mid-infrared spectra of this SNR available.

\section{Observations}

We mapped the vicinity of the Cas A SNR with the Heinrich Hertz Submillimeter Telescope (SMT) using the ALMA-type sideband separating 1.3 mm receiver on 2011 January 30.  We observed the J $=2-1$ transitions of $^{12}$CO and $^{13}$CO at 230 GHz and 220 GHz, simultaneously, configuring the receiver with $^{12}$CO J $=2-1$ in the upper sideband and $^{13}$CO J $=2-1$ in the lower sideband and using an IF of 5000 MHz.  The two sidebands were configured with independent mixers receiving both the horizontal and vertical polarizations.  We used two spectrometers in parallel, providing 1 MHz and 250 kHz filters.  The quality of the sky was good, with opacity around a few percent for the duration of the mapping.  The system noise temperature varied with source elevation over the course of the observation, with values of $\sim$ 200 K over the majority of the map, but values as large as 269 K over the last several rows.

The map was obtained on-the-fly (OTF) and covered a $10' \times 10'$ region centered on the SNR. The rows were spaced $10''$ in declination and the area was scanned boustrophedonically, with a line-free reference position observed after every other row. The final map has $61 \times 65$ pixels, each $10'' \times 10''$ with a beam width of $33''$ FWHM. Our reference position strategy provided adequately flat spectral baselines, although there was residual baseline ripple of $\sim$ 0.1 K peak-to-peak in some areas.

The SMT data were processed using CLASS.\footnote{\url{http://www.iram.fr/IRAMFR/GILDAS}}  A spectrum was extracted from each pixel and linear baseline subtraction was performed followed by sky subtraction using a spectrum derived from our reference.  The spectra were then gridded into a datacube and converted into MIRIAD format \citep{sault95} for analysis.

The infrared data are high-resolution $10-40$ $\mu$m spectra obtained in 2008 September (AORID 50322) using the \textit{Spitzer} IRS.  Two scans were constructed across the northern edge of Cas A consisting of 16 individual spectra in the short-high module integrating for 120 s per pointing.  Long-high spectra were obtained at the same pointings integrating for 60 s per spectrum.  The same parameters were used for each map, with pointing positions spaced approximately $2''$ apart (Figure 1).

All IRS data were reduced using SMART \citep{hig04}.  The two spectral scans were broken down into 16 individual spectra each for 32 total spectra at different pointings. Background subtraction was performed from sky images taken during the observation.  Finally, the spectra were extracted using the full-aperture algorithm for extended sources.

\begin{figure}
	\figurenum{2}
	\centering
		\includegraphics[width=0.4\textwidth]{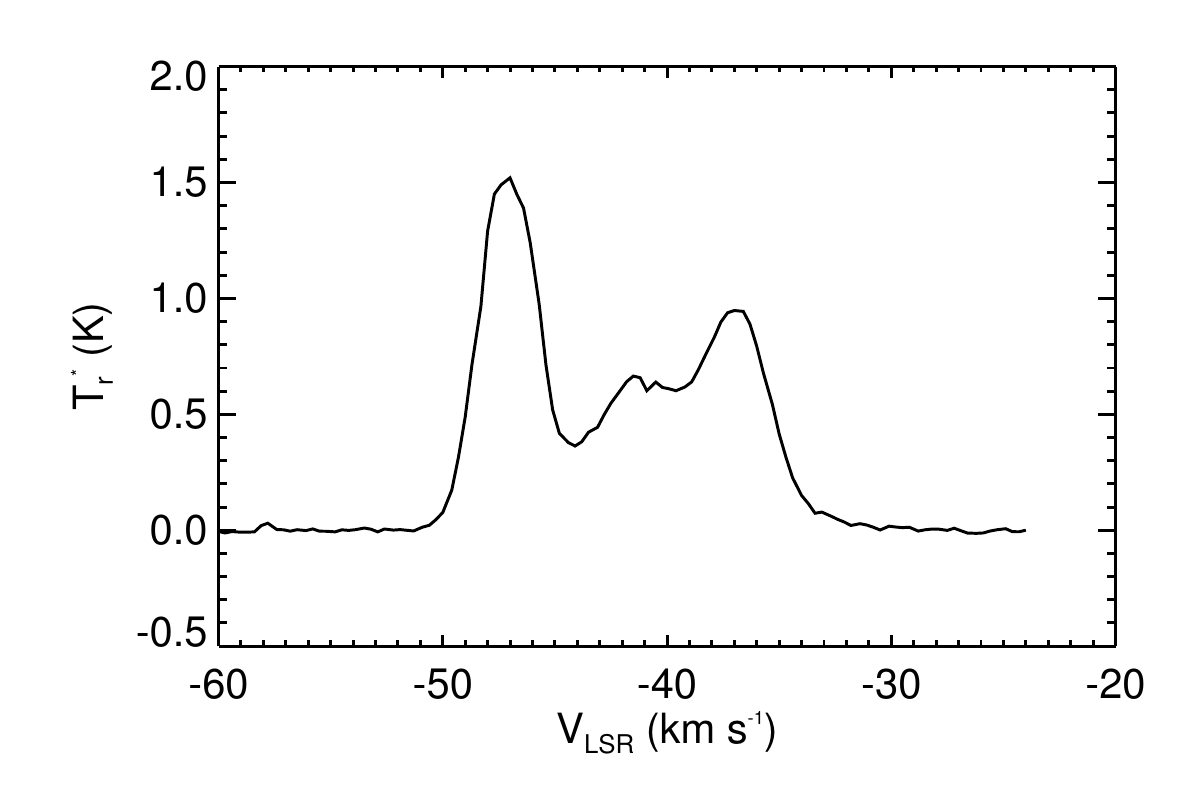}
	\caption{\scriptsize $^{12}$CO J=2-1, 250 kHz filter bank spectrum averaged across the $10' \times 10'$ mapped region.}
\end{figure}

\begin{figure*}
	\figurenum{3}
	\centering
		\includegraphics[width=0.4\textwidth,angle=270]{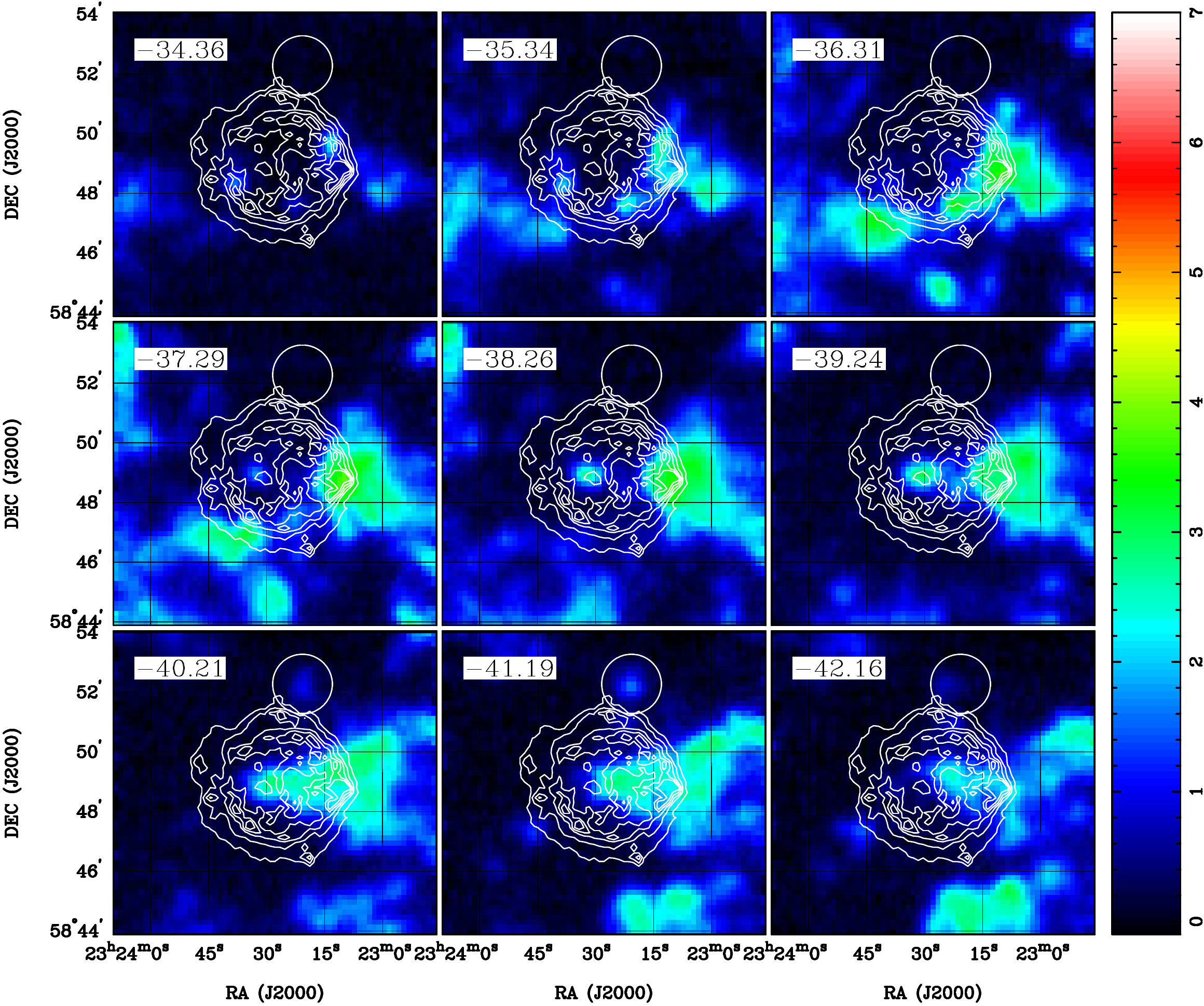}
		\includegraphics[width=0.4\textwidth,angle=270]{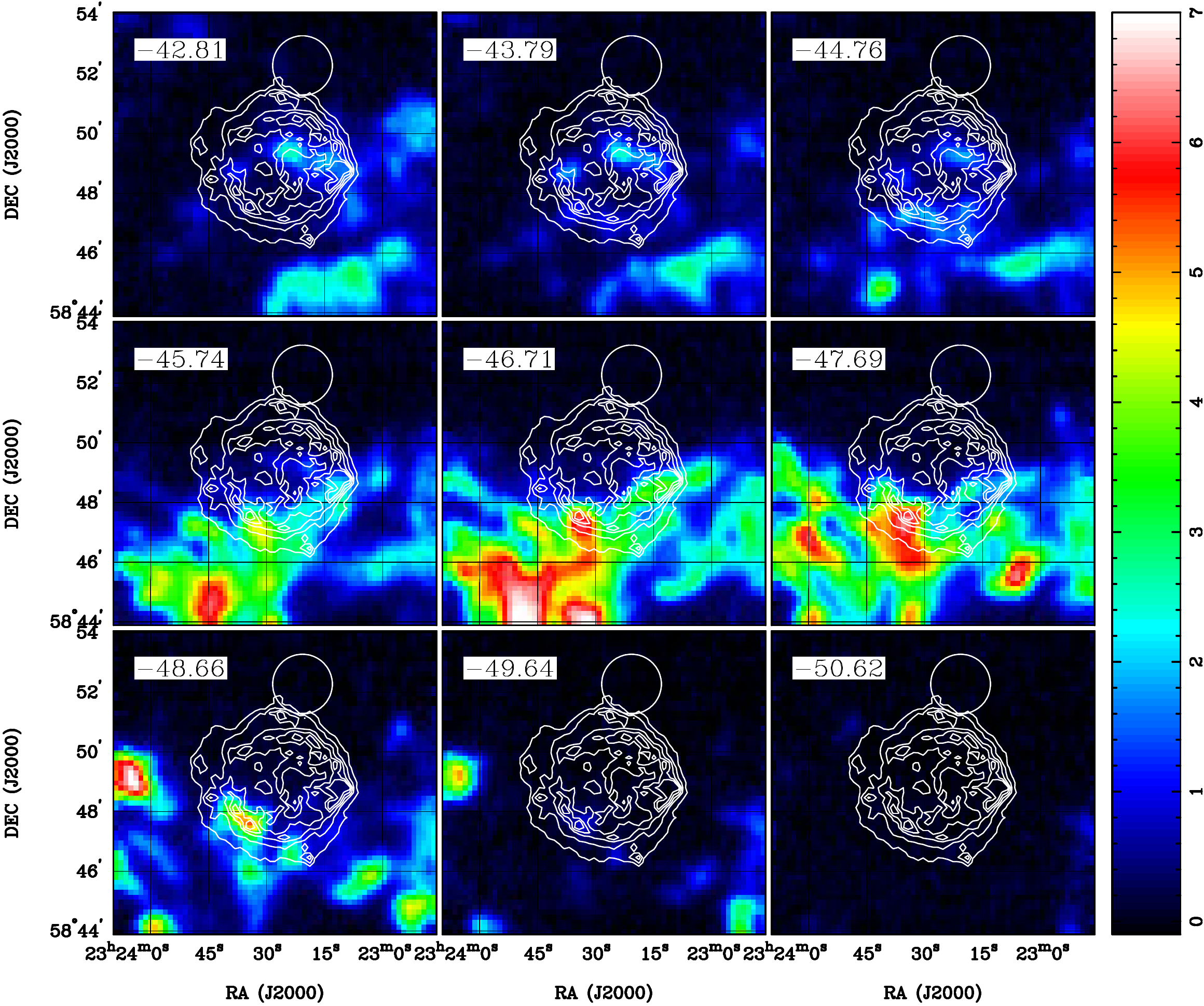}
		\caption{\scriptsize  Channel maps of $^{12}$CO J=2-1 in the 250 kHz filter with VLA 20 cm emission contour overlaid.  The velocity of each map is given in the upper left hand corner.  The circle depicts the location of an over-density of molecular gas located along the northern edge of the remnant.  Note that the over-densities of molecular gas to the west, south and east of the SNR correspond to peaks in the contours of 20 cm emission.}
\end{figure*}

\begin{figure}
	\figurenum{4}
	\centering
		\includegraphics[width=0.35\textwidth,angle=270]{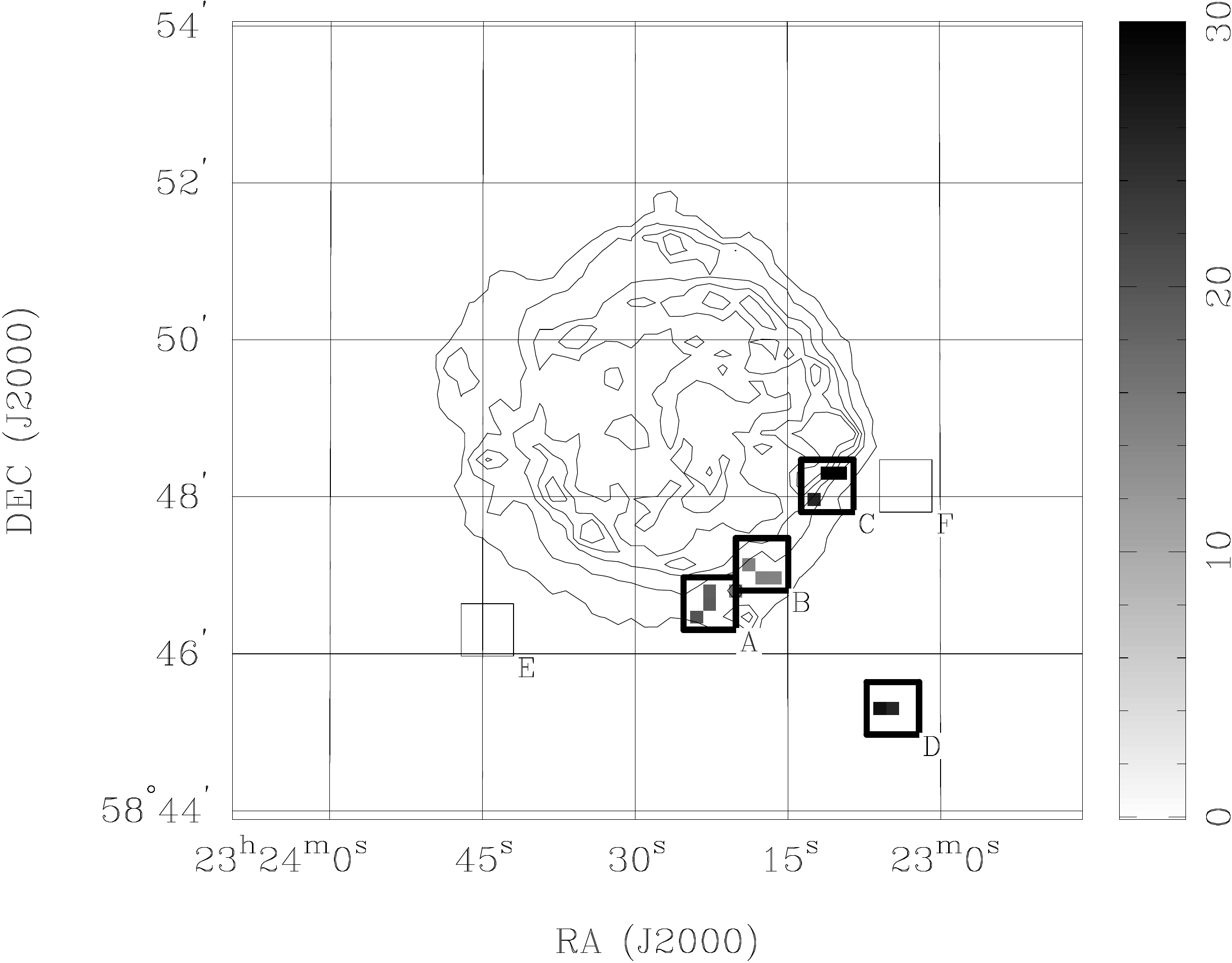}
	\caption{\scriptsize The integrated intensity (K km s$^{-1}$) of $^{12}$CO J=2-1, 250 kHz filter bank spectrum masked for regions where the FWHM of $^{12}$CO J=2-1 lines exceeds 6 km s$^{-1}$, which is a fiducial value we use to denote ``broad-line regions.'' The VLA 20 cm map of Cas A is overlaid to demonstrate the locations of various broad-line regions relative to the SNR.  The only four regions in the entire map dominated by broad lines (A, B, C, D) are indicated by shaded pixels within thick-lined boxes, while two empty, thin-lined boxes (E, F) are randomly selected comparison regions dominated by narrow lines.  The averaged spectra from each of these regions are plotted in Figure 5.}
\end{figure}

\section{CO Line Emission Around Cas A}

\subsection{Kinematics of CO Line Emission as Observed in J \texorpdfstring{$=2-1$}{} Spectra}

A spectrum of the $^{12}$CO J $=2-1$ transition observed using the 250 kHz filters and averaged over the entire $10' \times 10'$ mapped region is shown in Figure 2.  The high-velocity gas is differentiated into three main components as shown in the CO J $=1-0$ maps of \citet{ll99}.  The most highly blue-shifted gas ($-50$ to $-45$ km s$^{-1}$) is concentrated in filamentary structure to the south and southeast of the remnant, the intermediate velocity gas ($-45$ to $-39$ km s$^{-1}$) appears in the west and north edges, and the most red-shifted gas ($-39$ to $-34$ km s$^{-1}$) is located in the southeast and west.  These features can be observed in the channel maps in Figure 3 where the $^{12}$CO emission at 250 kHz resolution is displayed from $-33$ to $-50$ km s$^{-1}$.  Between $-33$ and $-41$ km s$^{-1}$, the brightest emission appears west and south of the SNR, with a separate cloud that extends to regions on the order of $1 - 2'$ away in projected radius along the line of sight of the remnant.

\subsection{Possible Cloud/SNR Interactions}

The frames at $-40.86$ and $-41.51$ km s$^{-1}$ in Figure 3 show the molecular cloud at the north edge of Cas A described previously \citep{hines04, ll99}.  The cloud is circled in white and lies near the two tracks of our \textit{Spitzer} IRS observations outlined in Figure 1.  The cloud has a mass of 650 M$_{\odot}$ \citep{hines04} and extends over a region $1.5' \times 0.8'$ on the sky.  Figure 1 also shows the enhanced zone of radio emission at the interface between the SNR and this cloud.  This over-density and the radio feature so close to the edge of the remnant suggest an interaction between the two media.

In general, the brightest CO-emitting regions along the line of sight to Cas A correspond well with the brightest emission in the VLA 20 cm map of \citet{arlpb91} possibly indicating additional interactions between the shock front and molecular clouds.

\subsection{CO Line Broadening}

At points of interaction where shocks from the supernova ejecta have propagated through the molecular gas, we would expect to see line broadening due to enhanced turbulence.  To evaluate the extent of that broadening in molecular clouds toward Cas A, we developed an algorithm to search the entire mapped area and velocity range in our 250 kHz filter maps for pixels with large line widths. The algorithm was optimized {\it not} to confuse multiple narrow lines with a single broad one. To this end, we identified pixels in our datacube with CO emission exceeding the noise baseline by at least 5-$\sigma$ and fit a Gaussian profile with a linear baseline to the line profile.  We iteratively increased the velocity range of interest around the velocity centroid of our fit until the fit either diverged (i.e., the reduced $\chi^{2}$ of the fit increased by a factor of two from the previous fit) or the fit parameters varied less than $10\%$ from the previous one.  In this way, we identified pixels that exhibit CO lines with widths exceeding $6$ km s$^{-1}$, our fiducial value for ``broad-line regions.'' We tested the sensitivity of the pixels selected on the threshold velocity width and found that the pixels identified as ``broad-line regions'' were not a strong function of the fiducial velocity-width above $\sim 3$ km s$^{-1}$.  That is, these pixels stand out as qualitatively distinct components of the cloud, with significantly higher line widths than the molecular emission in adjacent pixels.  Figure 4 locates the regions where the integrated CO intensity has strong wide components.  For comparison, we overlaid the VLA 20 cm continuum map to demonstrate where the broad-line regions lie in relation to Cas A.

In order to further examine the association between Cas A and these clouds, we have highlighted six regions in Figure 4 using boxes (A, B, C, D, E, F).  The first four of these regions contain pixels that our algorithm identified as having ``broad-line regions'' while the last two contain only narrow lines in a similar velocity range for comparison.  The average spectra of the pixels in each of the six boxes are plotted in Figure 5.  The blue lines in panels A and B and the red lines in panels C and D do indeed appear broad, with FWHM velocities exceeding $6$ km s$^{-1}$ (see Table 1 for values from our Gaussian fits to these lines).  Furthermore, these lines appear to demonstrate the ``broad-wing'' characteristic that is a signature of shocked molecular emission toward SNRs \citep[e.g.,][]{den79a, den79b, burt88}.  By comparison, the molecular emission lines in panels E and F appear narrower, which we approximate as Gaussian profiles with FWHMs less than $3$ km s$^{-1}$.

\begin{deluxetable}
{ccccc}
\tabletypesize{\scriptsize}
\tablecaption{Gaussian fit values for CO lines}
\tablewidth{0pt}
\tablehead{
Line & Peak (K) & v$_{c}$ (km s$^{-1}$) & FWHM (km s$^{-1}$) & rms (K)
}
\startdata
A$_{blue}$ & 3.12 & $-46.9$ & 6.48 & 0.440\\
A$_{red}$ & 0.73 & $-37.2$ & 3.42 & 0.083\\
B$_{blue}$ & 2.09 & $-46.5$ & 6.45 & 0.158\\
B$_{red}$ & 0.81 & $-36.4$ & 2.74 & 0.061\\
C$_{blue}$ & 1.40 & $-46.5$ & 3.19 & 0.051\\
C$_{red}$ & 3.15 & $-37.7$ & 6.19 & 0.116\\
D$_{blue}$ & 1.40 & $-46.6$ & 3.01 & 0.051\\
D$_{red}$ & 3.55 & $-37.7$ & 6.66 & 0.097\\
E$_{blue}$ & 3.99 & $-46.8$ & 2.71 & -0.009\\
E$_{red}$ & 2.51 & $-36.5$ & 2.78 & 0.022\\
F$_{blue}$ & 2.40 & $-48.5$ & 2.50 & 0.005\\
F$_{red}$ & 0.42 & $-37.3$ & 2.69 & 0.027\\
\enddata
\end{deluxetable}

We estimated the masses of the clouds located west, south, and southeast of the SNR, and the momentum necessary to generate the broad lines in these regions.  We set a threshold of $6$ km s$^{-1}$ in the second moment of $^{12}$CO emission at each pixel to define the turbulent regions.  For clouds with observed sizes on the order of a few parsecs, this value is slightly more than twice as large as $^{12}$CO line widths typically observed in quiescent galactic molecular clouds \citep{hb04}. Using the ratio of $^{12}$CO to $^{13}$CO emission, we calculated the optical depth toward each pixel in three of the molecular clouds and used this value to determine the column density of molecular hydrogen at each position.  From these values, we found cloud masses of approximately $660, 530,$ and $340$ M$_{\odot}$ for the turbulent regions of the western, southern, and southeastern clouds, respectively.  The amount of momentum necessary to generate these line widths is then approximately, M$_{\mbox{\tiny cloud}} \sigma_{\mbox{\tiny CO}}$ or $4000, 3200, 2000$ M$_{\odot}$ km s$^{-1}$ assuming a minimum line width of $6$ km s$^{-1}$ for each cloud.

\begin{figure}
	\figurenum{5}
	\centering
		\includegraphics[width=0.4\textwidth]{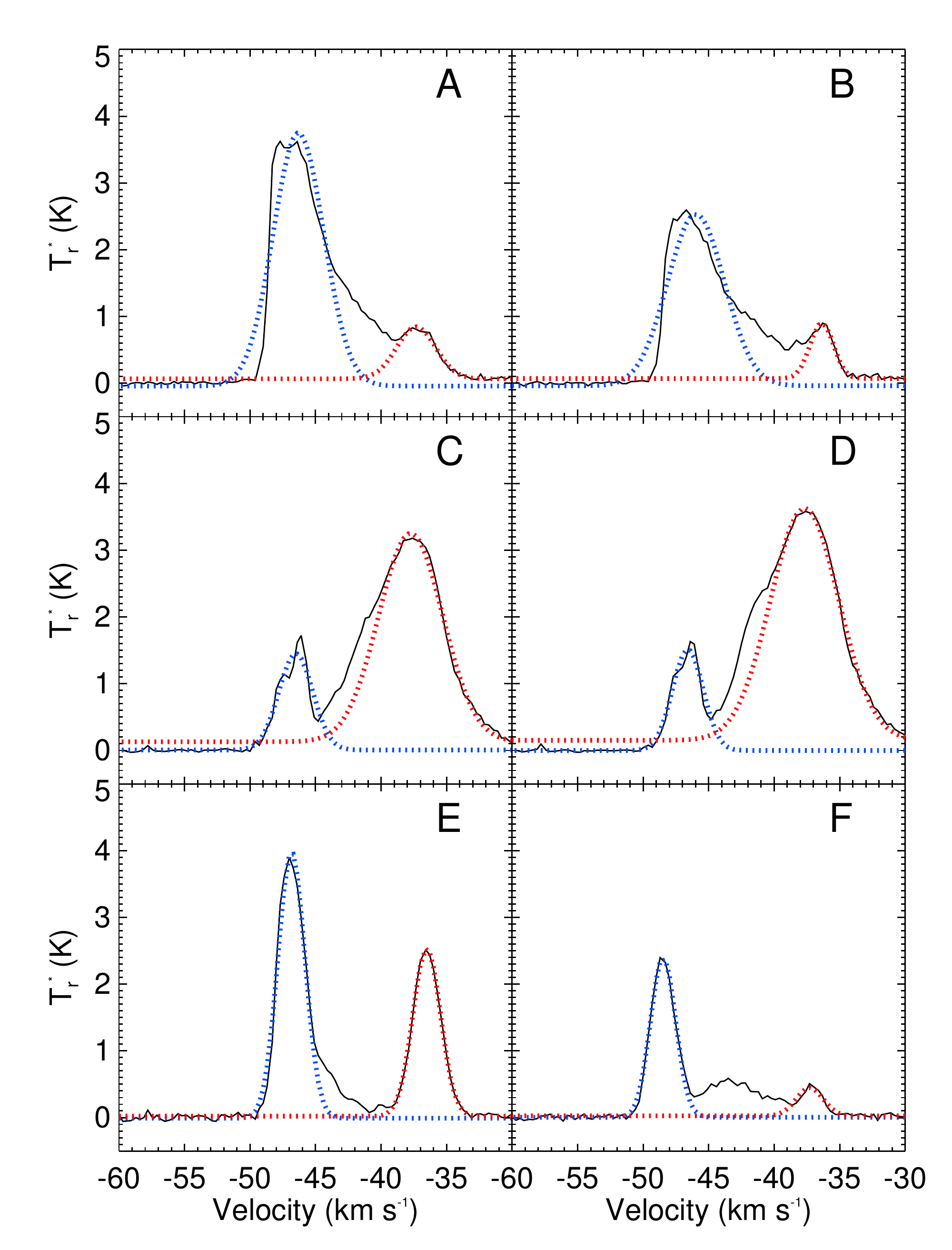}
	\caption{\scriptsize $^{12}$CO J=2-1, 250 kHz filter bank spectrum averaged across the six roughly $50'' \times 50''$ regions denoted in Figure 4.  Note that each spectrum contains multiple CO line components, although we have fitted Gaussian curves to the two brightest components in each panel.  In each panel, we call the feature below $\sim -43$ km s$^{-1}$ the ``blue'' component and the feature above $\sim 40$ km s$^{-1}$ the ``red'' component, denoted by the blue and red dotted curves.  The blue components in panels A and B and the red components in panels C and D are broadened, that is, they have a FWHM exceeding $6$ km s$^{-1}$ (line fits given in Table 1).  We compare these lines to the narrower CO lines in panels E and F, which are more typical of CO in the ISM.}
\end{figure}

\begin{figure*}
	\figurenum{6}
	\centering
		\includegraphics[width=0.46\textwidth]{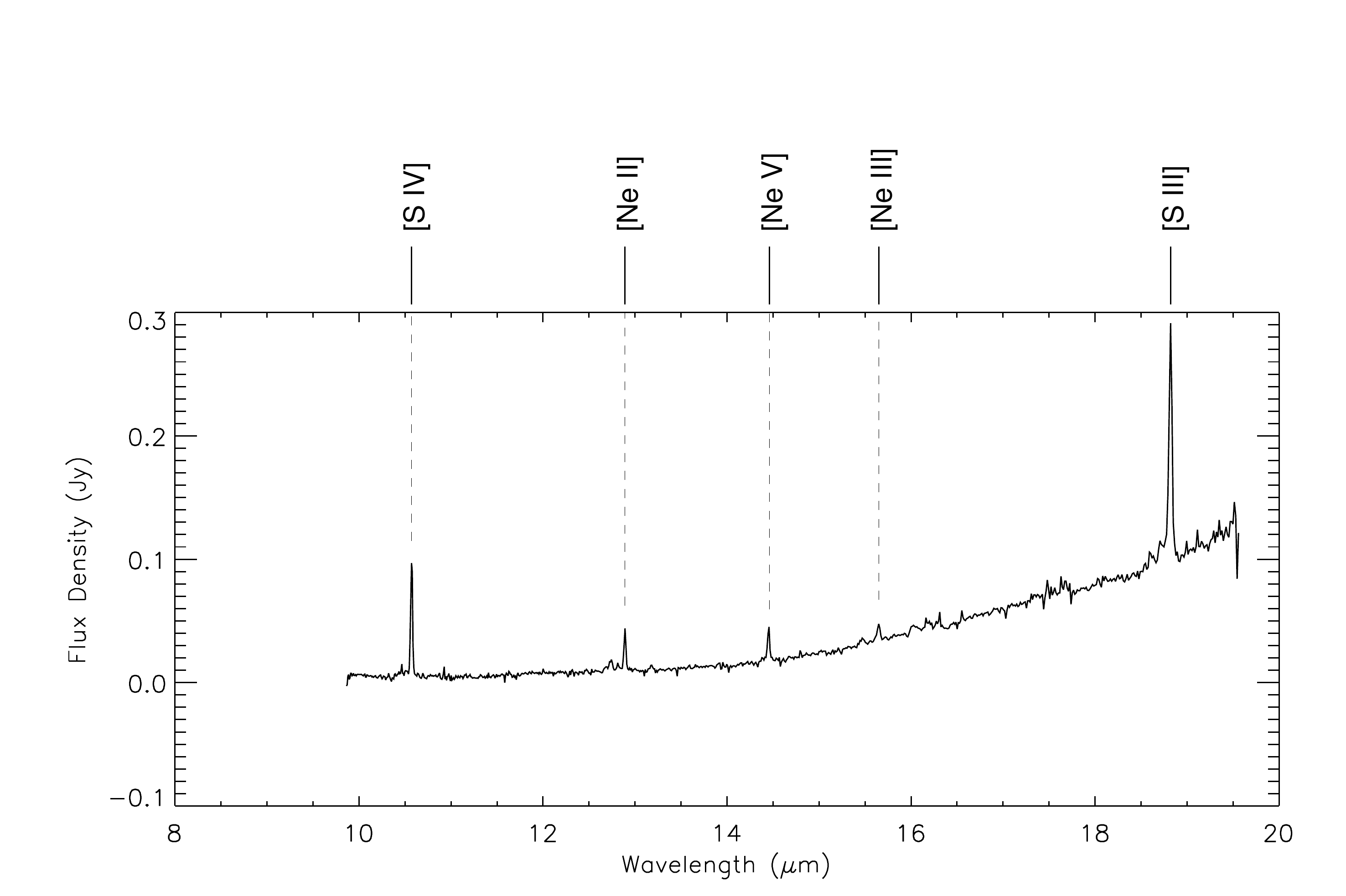}
		\includegraphics[width=0.46\textwidth]{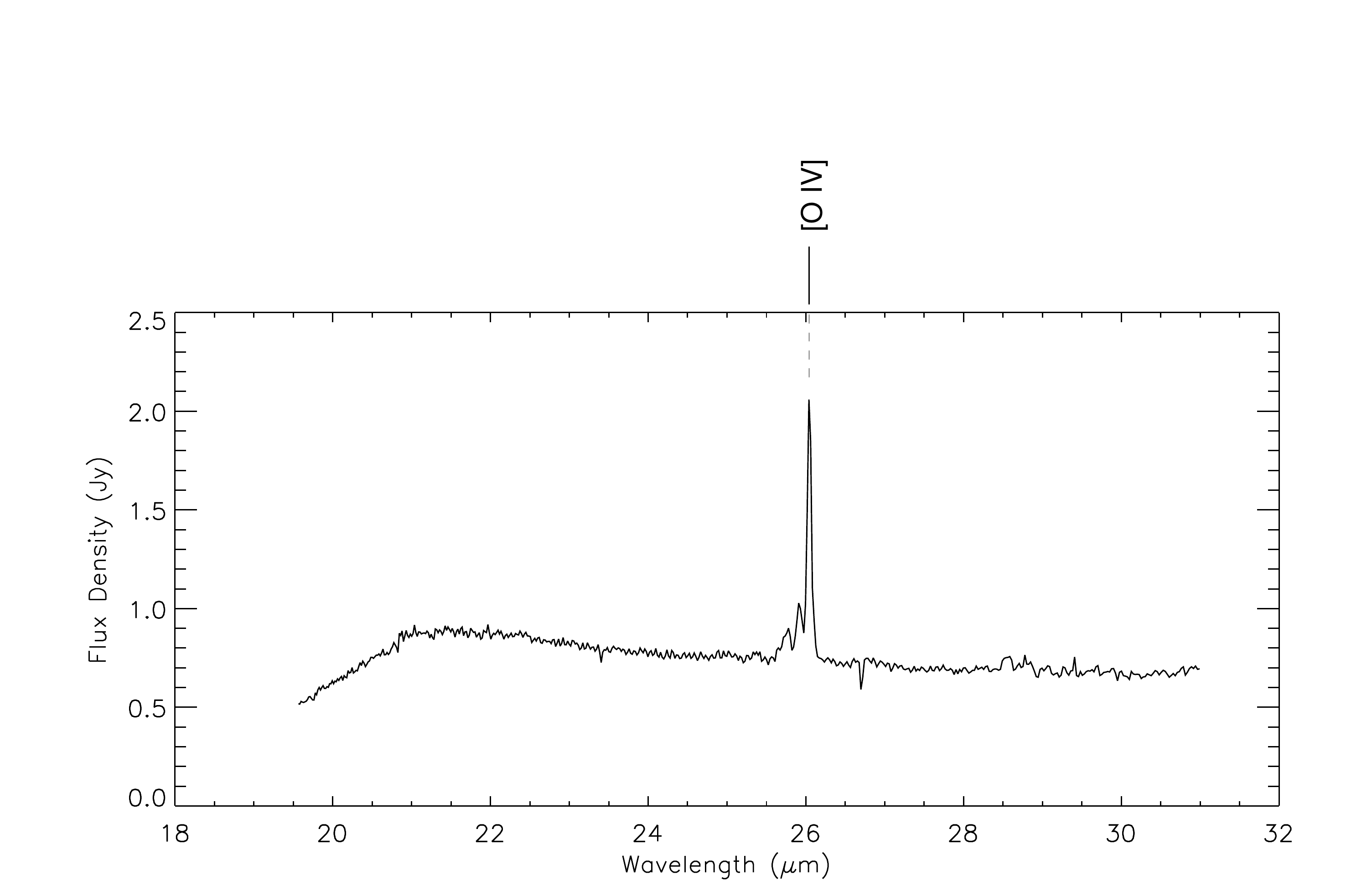}
	\caption{\scriptsize Example of high-resolution shock front spectrum from the \textit{Spitzer} IRS high-resolution data.  This pointing represents the southernmost pointing along scan 2 (west in fig. 1).  The spectrum is separated into the short-wavelength (SH - upper panel) and long-wavelength (LH - lower panel) modules.  All spectral features observed in each scan are identified.}
\end{figure*}

Analysis of infrared transitions of CO toward Cas A suggests that astrochemical processes form molecular gas in the ejecta-rich reverse shock of the remnant \citep{rho09, rho12}.  Indeed, \textit{Herschel} detections of broad, high-J rotational CO lines in the far infrared toward the Cas A remnant suggest that molecular gas has reformed recently in bright knots\citep{wall13}, reminiscent of optical fast moving knots \citep{fes01b}. However, the masses we detect in the western, southern, and southeastern clouds greatly exceed the mass of the progenitor star, so we infer that the molecular gas cannot originate from the remnant itself.

Some of the broad-line regions (A, B, C) we have highlighted lie not merely toward the edge of the Cas A remnant, but toward radio continuum enhancements along the western and southern edges of the remnant, possibly due to an interaction between molecular gas and the supernova ejecta.  The western position matches the filamentary structure observed in previous work \citep{ll99} and lies where the molecular cloud morphology is consistent with an interaction as previously suggested by \citet{kra96}.  However, one of the broad line regions (D) appears more than $2'$ from the forward shock, as demarcated by the edge of the remnant, which suggests an interaction with very fast-moving ejecta. We will explore this behavior in Section 5.1.

\section{Analysis of Cas A Infrared Spectroscopy}

\subsection{Spectral Features}

A typical infrared spectrum with spectral line identifications, taken from one of our pointings toward the northern Cas A shock front, is shown in Figure 6.  Both the emission lines and the dust continuum emission become less bright as the spectra are taken from pointings that extend further north of the remnant.  In our example spectrum, the continuum is relatively faint at shorter wavelengths but peaks strongly around 21 $\mu$m and slowly decreases in brightness at longer wavelengths.  This 21 $\mu$m peak observed by \citet{lagage96}; \citet{adm99}; and \citet{rho08} has been associated with freshly formed silicate and iron dust in the ejecta.

In scan 1, emission features are apparent at 10.5, 12.8 (doublet), 14.3, 18.7, and 25.9 $\mu$m.  Similarly, scan 2 presents several strong features in the pointings furthest south.  These were identified at 10.5, 12.8 (doublet), 14.3, 15.6, 18.7, and 25.9 $\mu$m.  Virtually all the features identified in scans 1 and 2 correspond with previously identified fine-structure lines based on low-resolution IRS spectra \citep{rho08, smith09, del10}.  In particular, [S IV] (10.51 $\mu$m), [Ne II] (12.81 $\mu$m), [Ne V] (14.32 $\mu$m), [Ne III] (15.56 $\mu$m), [S III] (18.71 $\mu$m), and [O IV]+[Fe II] (25.94 $\mu$m) are commonly observed features in the ejecta toward this remnant.  Apart from these lines, there is also a second emission line near 12.7 $\mu$m in both scans 1 and 2.  Given the high-velocity components in other spectral features, we consider in the appendix whether this line can originate from neon in high-velocity gas.

The bright features (e.g., [O IV], [S III], [S IV]) also show considerable velocity structure.  We have plotted these features in velocity-space for a spectrum composed of the two southernmost pointings along scan 2 averaged together (Figure 7).  These averaged spectra emphasize the high-velocity features, which we have marked with vertical lines.  The remainder of our analysis concentrates on the [O IV] line; the other lines are discussed in the appendix.

\begin{figure}
	\figurenum{7}
	\centering
		\includegraphics[width=0.45\textwidth]{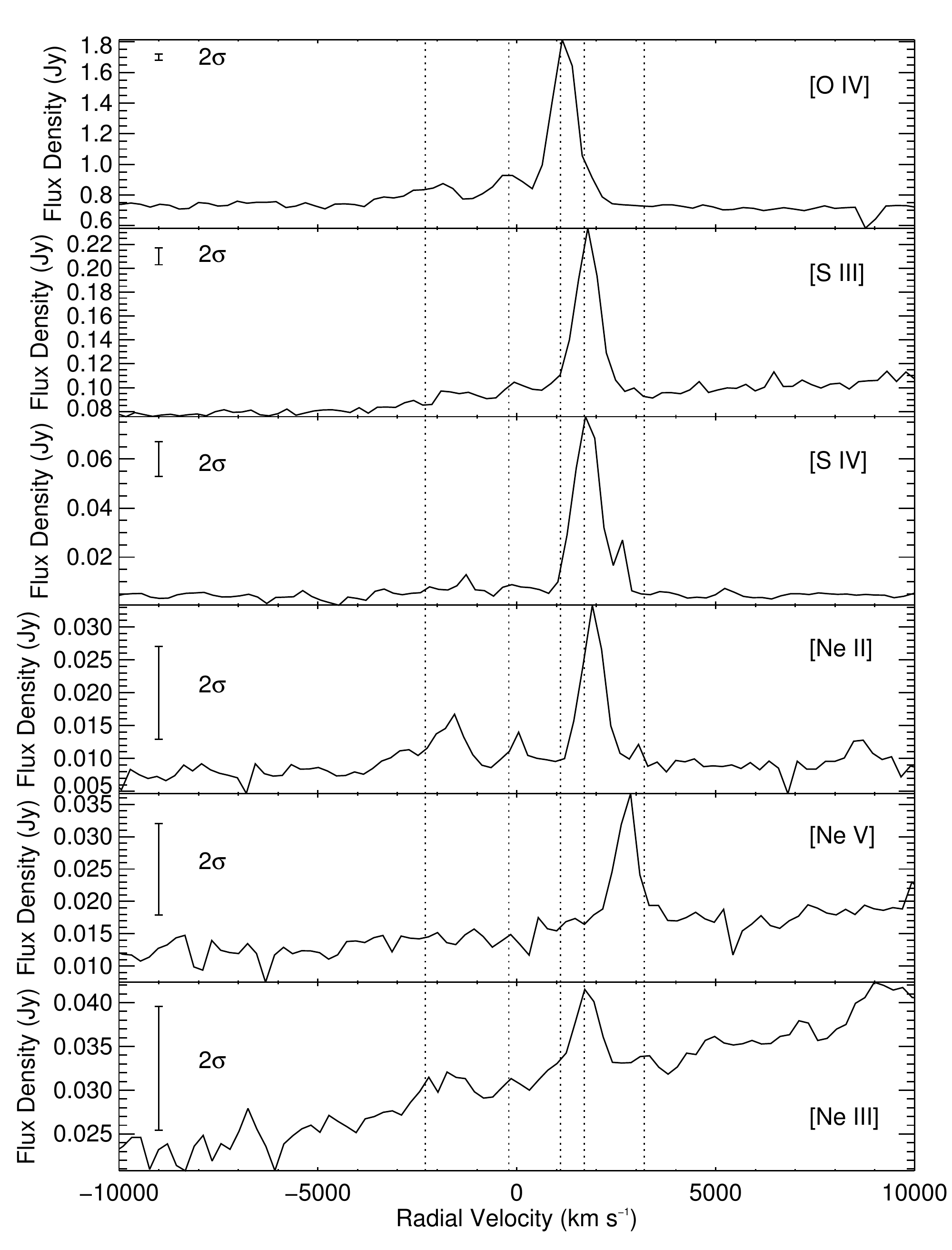}
	\caption{\scriptsize Spectra from the six ([O IV], [S III], [S IV], [Ne II], [Ne V], [Ne III]) fine structure lines identified in Figure 6.  These spectra represent the two southernmost pointings of scan 2 averaged to emphasize the presence of high-velocity features.  A sample 2-$\sigma$ error bar is shown in the upper-left hand corner to compare with the strength of the identified features.  Five dotted, vertical lines mark some recurrent high-velocity features (at $-2300, -200, 1100, 1700, 3200$ km s$^{-1}$).}
\end{figure}

\begin{figure}
	\figurenum{8}
	\centering
		\includegraphics[width=0.45\textwidth]{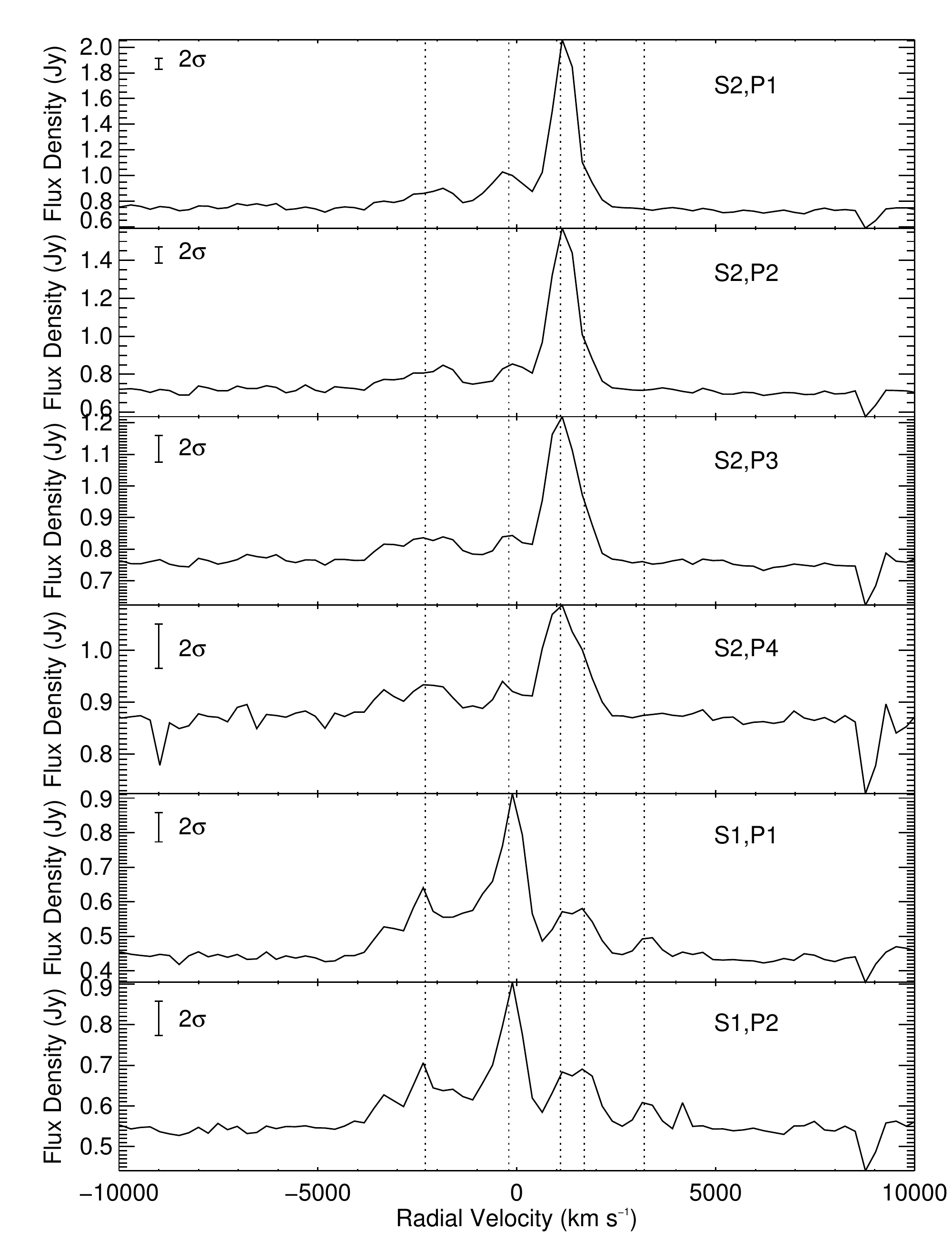}
	\caption{\scriptsize Identified [O IV] (25.94 $\mu$m) lines in scans 1 and 2 (S1 and S2) at the southernmost pointings in each scan (P1-2 in S1, P1-4 in S2).  These lines have been plotted versus relative velocity to demonstrate the presence of high-velocity components in the spectra.  A sample error bar is shown in the upper-left hand corner to compare with the strength of the identified features.  Five dotted, vertical lines mark some recurrent high-velocity features (at $-2300, -200, 1100, 1700, 3200$ km s$^{-1}$).}	
\end{figure}

\subsection{[O IV] 25.94 \texorpdfstring{$\mu$}{}m line}

\subsubsection{Observations}

The [O IV]+[Fe II] feature around 25.94 $\mu$m is consistently the brightest fine-structure line detected toward the Cas A remnant.  It is brightest around the edge of the remnant with significant emission toward the center relative to other lines \citep{rho08}.  From the co-added spectra in both scans, it is possible to separate high-velocity features of the supernova outflow in the resolved [O IV] line profiles (around 25.94 $\mu$m).  At pointing 1, there are five features centered at 25.63, 25.74, 25.86, 26.14, and 26.23 $\mu$m (Figure 8).  Assuming most of the flux is due to [O IV] emission, these wavelengths correspond to radial velocities of $-3580$, $-2300$, $-920$, $2310$, and $3350$ km s$^{-1}$, in agreement with the range of previous velocities observed within the remnant ($-4000$ to $6000$ km s$^{-1}$ \citep{del10}).

Along scan 2, there appear to be three distinct features in the stacked spectrum at wavelengths 25.76, 25.92, 26.14 $\mu$m.  For [O IV] emission, these wavelengths correspond to radial velocities of $-2080$, $-230$, and $2310$ km s$^{-1}$.  At the high signal to noise on the central features in the scan 1 and 2 spectra, the wavelength difference between them (1.5 full IRS line widths) is significant. Given the modest spectral resolution (R $\sim$ 600) and signal to noise, there is no convincing case that any of the features coincide between scans 1 and 2 (see Figure 8).  

In any case, the two spectra demonstrate that we observe ejecta over a velocity range of at least 5000 km s$^{-1}$ in a relatively small region of the northern shock front (see Figure 1). To demonstrate the behavior of [O IV] over the spatial region for which we have high-resolution spectra, we have plotted spectra for all 32 pointings (see description in Section 2) centered around the [O IV]+[Fe II] line in Figure 9.  A linear baseline was subtracted and flux was added to each spectrum such that the line with the brightest baseline corresponds to the pointing furthest south (i.e. P1) and each subsequent spectrum stepping down in flux represents the next pointing spaced $2''$ further north (i.e. P2, P3, etc.).  We note in particular that the bright [O IV] feature disappears over an angular distance of about $6''$ for S1 and $10''$ for S2, which is comparable to the IRS slit width of $11''$ \citep{houck04}, and the relative emission from each high-velocity component remains the same at each pointing within the signal-to-noise.  That is, the 5000 km s$^{-1}$ velocity range persists all the way to a sharp cutoff in the line emission with increasing radius.

\begin{figure}
	\figurenum{9}
	\centering
		\includegraphics[width=0.23\textwidth]{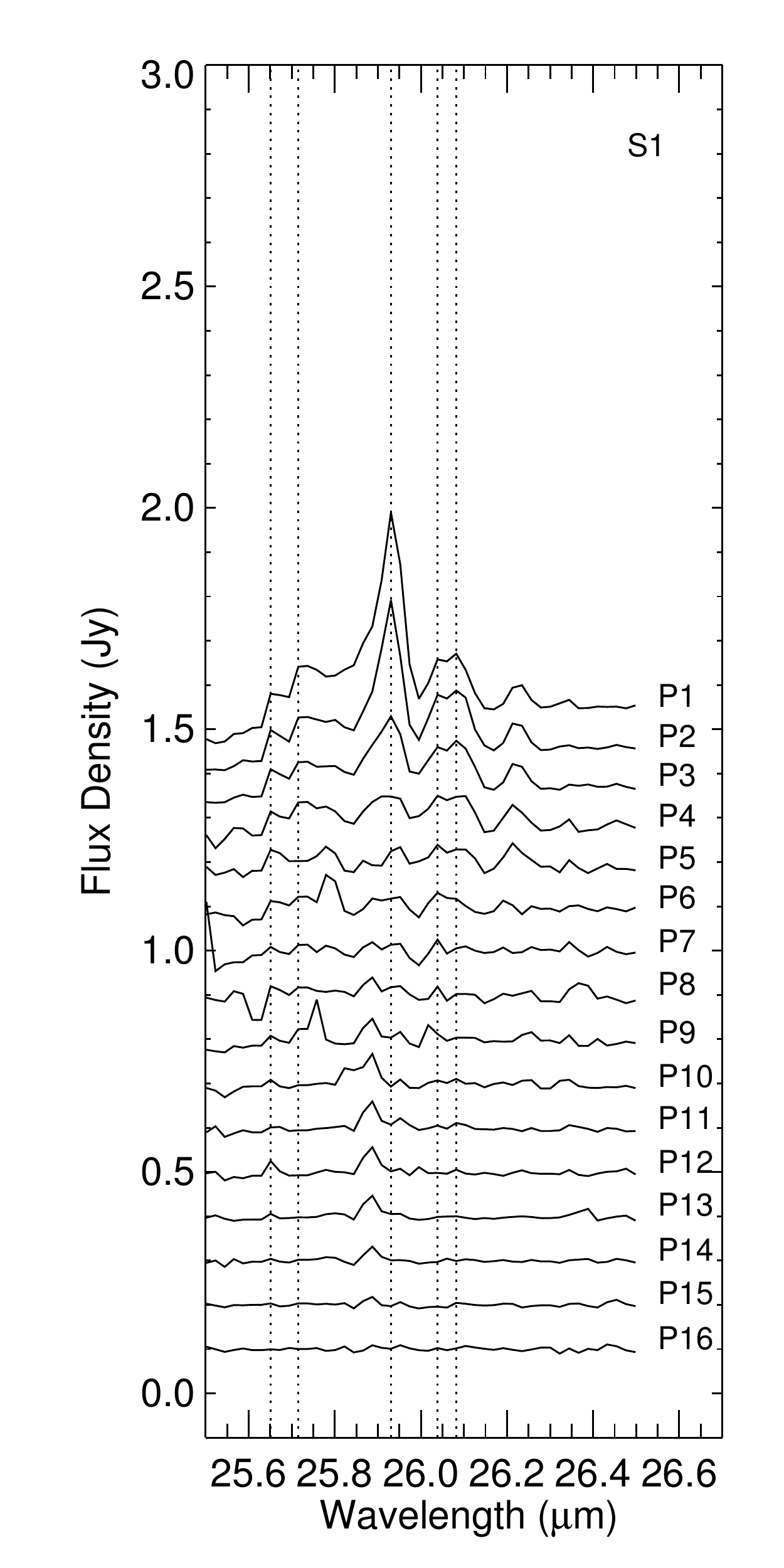}
		\includegraphics[width=0.23\textwidth]{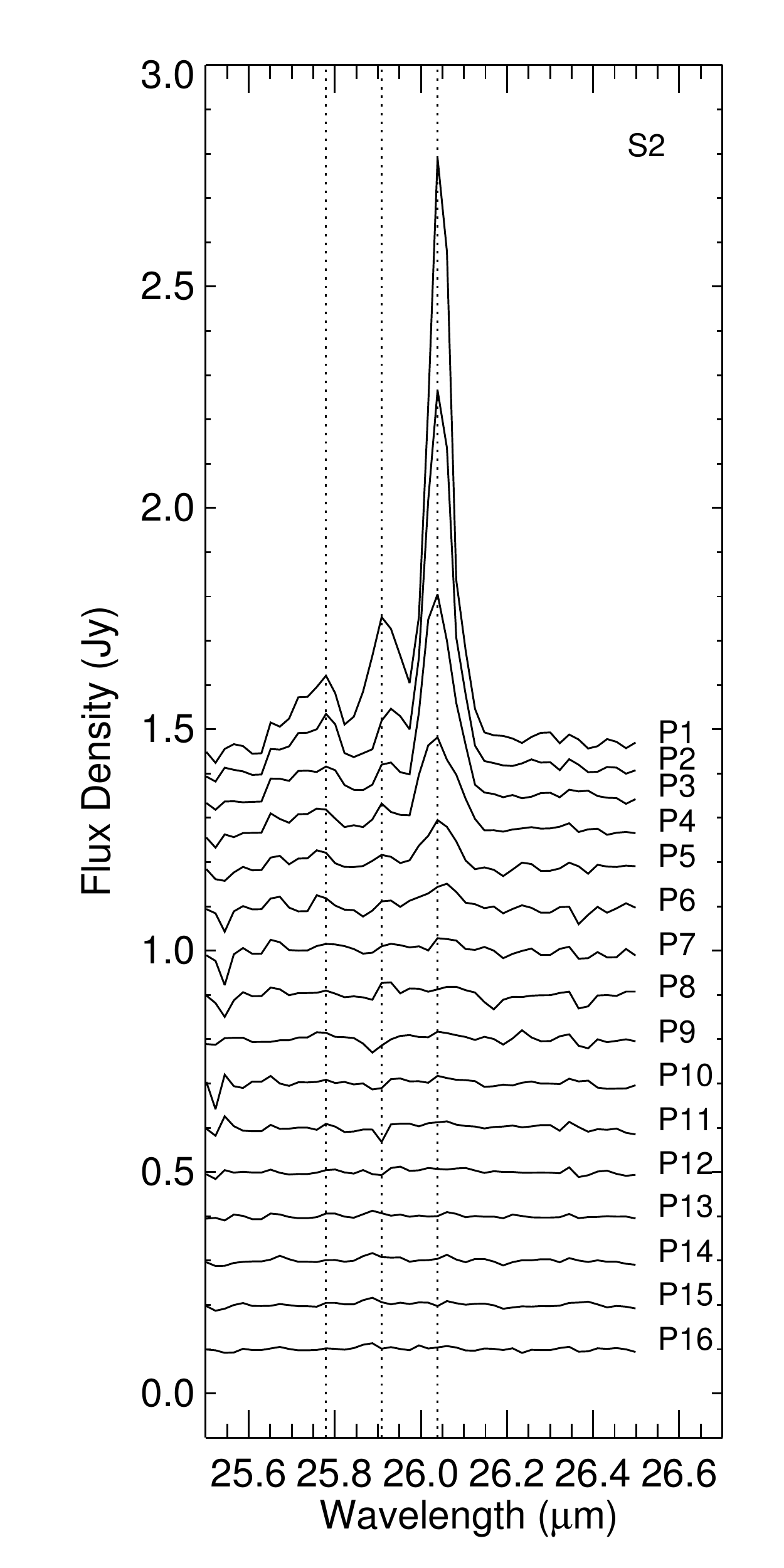}
	\caption{\scriptsize [O IV] 25.9 $\mu$m spectra depicted across the shock front along two scans, given as S1 (left) and S2 (right), to the north of Cassiopeia A.  The spectra are depicted with the southernmost pointing (closest to Cas A center) at the highest flux density with spectra at pointings successively further north at lower flux density.  These pointings start at a position approximately $3'$ north of the center of Cas A and step north in increments of $2''$.  Note that the relative flux of high-velocity features surrounding the [O IV] line (dotted lines at $-3580$, $-2300$, $-110$, $1140$, and $1640$ km s$^{-1}$ in S1 and $-1850$, $-360$, and $1140$ for S2) remain roughly the same at each pointing and the feature disappears completely after roughly $3$ pointings, or $6"$, for S1 and $5$ pointings, or $10"$, for S2, which is comparable to the slit width of \textit{Spitzer} IRS ($11''$).  This feature suggests that there is a sharp cut off where the [O IV] emission, and thus the supernova ejecta, rapidly becomes less dense, consistent with a contact discontinuity between the ejecta and interstellar gas.}
\end{figure}

\subsubsection{Is the velocity spread due to turbulence in the supernova ejecta?}

The velocity spread in the infrared fine structure lines is similar to that observed over most of the face of the SNR, where it is attributed to two causes: 1) turbulence in the explosion ejecta; and 2) Doppler shifts from systematic motions associated with radial expansion of the remnant. \citet{del10} show that interpreting the extreme red and blue line components as due to Doppler shifts associated with line emission from opposite sides of an expanding sphere can yield a three-dimensional picture of the remnant. In this picture, the absolute Doppler shifts should tend to a minimum as one approaches the extreme edge of the remnant as projected onto the sky, since all of the systematic motions associated with the spherical expansion will lie on the plane of the sky. We will analyze this behavior quantitatively in the following section.

Does the broad velocity spread in our observations, obtained at the extreme edge of the remnant, have the same causes? Attributing it to a combination of turbulence in the ejecta and the systematic expansion of the remnant has a number of implications. It might indicate that the lines we observe do not come from the extreme edge of the remnant, although we show in the next section that this explanation is not likely. It might indicate that the spherical geometry is not strictly correct at the edge of the remnant. Broadened lines around the extreme edge of the remnant might arise if the remnant were like a cylinder which we see end-on, but such a model seems physically untenable.

As an alternative, in the following discussion we explore whether the turbulence is established in the area of our observations, which by Newton's second law requires that there be mass for the ejecta to react against to modify their outward expansion.  

\subsubsection{Does Radial Expansion Account for the Velocity Spread?}

We now test whether the large spread in radial velocities discussed above can be explained in the context of pure radial expansion of the SNR. Assuming that the observed emission originates from ejecta in a thin shell toward the Bright Ring of Cas A, the observed Doppler velocity ($v_{D}$) of ejecta in that shell has been modeled using [Ar II] (6.99 $\mu$m) in \citet{del10} using a three-parameter model,

\begin{equation}
w = |v_{D} - v_{c}| = \sqrt{(v_{c} - v_{m})^{2} - (r_{P}/S)^{2}}
\end{equation}

\noindent where $r_{P}$ is the projected radius in arcseconds, $v_{c}$ represents the center of the distribution of Doppler velocities, $v_{m}$ represents the lowest Doppler velocity for ejecta in the shell, $S$ is a scale factor that relates Doppler velocity to projected radius, and $w$ is the spread in velocities around $v_{c}$.  We note that as $r_{P}$ increases, $v_{D}$ approaches $v_{c}$ and the line width tends to zero.  This behavior is further expressed in the derivative of equation (1),

 \begin{equation}
 dw = \frac{2 r_{P} dr_{P}}{S^{2} \sqrt{(v_{c} - v_{m})^2 - (r_{P}/S)^2}}
 \end{equation}
 
For the parameters that \citet{del10} derive toward the Bright Ring of [Ar II] ($v_{c} = 859 \pm 100$ km s$^{-1}$, $v_{m} = -4077 \pm 200$ km s$^{-1}$, $S = 0.022'' \pm 0.002''$ per km s$^{-1}$), for projected radii close to the furthest extent of the Bright Ring of [Ar II] ($r_{P} \sim 100''$) and assuming incremental steps away from this position of $dr_{P} \sim 2''$, we find that $dv_{D} \sim 200$ km s$^{-1}$.  That is, for each $2''$ step in projected radius away from the center of expansion, $w$ should decrease by about $400$ km s$^{-1}$, or about $2000$ km s$^{-1}$ over all five pointings in which we see the lines.  In each spectral scan, the line components remain at the same relative strength until the ejecta are no longer present in our spectra. The observed spread in Doppler velocities of the [O IV] line in Figure 9, roughly a few $1000$ km s$^{-1}$, cannot be reconciled with the $129''$ projected radius of the spectral pointings from the observed center of expansion.  We infer that there must be some additional source of energy that drives the observed velocity spread.  

Finally, any explanation of the high-velocity Doppler components of the [O IV] emission must account for a range in velocities that is significantly in excess of the velocity spread typically observed in Cas A ejecta at a single position \citep[Figure 9 in][]{del10}.  This aspect of ejecta in the Cas A remnant has been confirmed through observation of optical ejecta knots at projected radii $2 - 3'$ and with Doppler velocities extending on the order of a few $100$ km s$^{-1}$ per knot \citep{fes01b}. \citet{isensee12} find that, to first order, the FWHM of the [O IV] line averaged over the entire remnant in low-resolution \textit{Spitzer} IRS spectra is $\sim 1000$ km s$^{-1}$.  We contrast these observations with the [O IV] emission at one position, at $130''$ projected radius from the expansion center of Cas A where high-velocity emission appears to extend $5000$ km s$^{-1}$.  Any explanation of the ejecta at this position must account for such an anomalously large spread in Doppler velocity.

\subsubsection{Fast Moving Knots and the Velocity Spread}

We now consider the possibility that the emission results from a superposition of multiple fast-moving knots. Optical studies of regions beyond the bright, nebular shell surrounding Cas A have revealed dozens of emission-line knots at projected radii $\sim 130 - 210$ arcseconds from the estimated center of expansion \citep{fes01, hf08}.  These observations cover the radius at which we observe highly Doppler shifted [O IV]+[Fe II] emission at approximately $23^{\mbox{h}}23^{\mbox{m}}26^{\mbox{s}}$, $58^{\circ} 50' 57''$, or about $130''$ north of the center of expansion as determined in \citet{thor01} at a position angle of roughly $354^{\circ}$.  However, catalogs of fast-moving knots do not show any in the vicinity of our observations \citep{hf08}.

Fast-moving knots tend to have a limited spread in velocity and exhibit structure on relatively small scales, with typical knot sizes between 0.2'' and 0.4'' \citep{fes01, del10}.  We observe multiple high-velocity Doppler components in each of our pointings, which would indicate that {\it several} fast-moving knots are required to account for the line structure. However, each pointing (through, roughly, P5 in Figure 9) exhibits emission at the same velocities.  This observation suggests that we observe a single line-emitting region with a broad range of velocities that sharply cuts off radially within approximately one slit width.

The presence of multiple FMKs that could mimic a single emitting region within spatial scales of $10''$ is unlikely in this part of the Cas A remnant.  The majority of known FMKs concentrate in the region along the northeast to southwest axis of the remnant where the ``jet'' is observed.  Along the northern rim, ejecta knots are significantly less common.  \citet[see Figure 10, therein]{hf08} find that there are 36 knots with projected radii between $130''$ and $140''$, or approximately $15$ per square arcminute.  In the on-sky area covered by the IRS slit ($\sim 250$ square arcseconds in Long-High mode), we would expect to observe approximately one FMK, or two FMKs in two separate pointings.  For a Poisson random distribution of FMKs on-sky, we find that the probability of observing $8$ FMKs in our two pointings (to match the eight velocity components in [OIV]) is roughly $0.1\%$.  Since no knots are observed at the slit positions, this low probability is confirmed empirically.

\subsubsection{Interaction-Induced Turbulence and the Velocity Spread}

We find that a scenario involving turbulent acceleration of the ejecta due to an interaction with interstellar material matches the observed spectra much more fully.  The large spread in Doppler velocities over this small region, the lack of any known ejecta knots in this region, and the placement of this region beyond the observed Bright Ring of optical emission support the interpretation of ejecta located toward the forward shock interacting with interstellar gas.  However, for a more in-depth analysis of the [O IV]+[Fe II] emission, we need to extend the study beyond the limited region covered by our high-resolution spectra (see Figure 1).

\subsection{Analysis of Low-Resolution [O IV] Emission}

We use archival low-resolution mid-infrared spectra that cover the remnant in its entirety to observe other regions along the shock front.  Based on the range of velocities in [O IV] in our \textit{Spitzer} IRS measurements, we expect that low-resolution data will exhibit a broadened spectral line indicating the spatial extent of Doppler shifted [O IV] emission in the SNR.  We therefore obtained low-resolution \textit{Spitzer} IRS spectra of Cas A \citep{rho08, smith09, del10} from the Spitzer Heritage Archive (AORID 3310) to compare the line width of [O IV] across the SNR shock front.

As a test, we convolved our own high-resolution data with a spectral response function for the IRS.  We derived this function by fitting a Gaussian curve to a narrow [O IV] emission feature observed in a planetary nebula (PNG 002.7-04.8) using the low-resolution module on \textit{Spitzer} IRS and obtained from the Spitzer Heritage Archive (Figure 10).  The convolution of our data to a lower spectral resolution effectively smooths and broadens the separate [O IV] features to a single line component.  The IRS low-resolution Cas A spectra will resemble this convolved spectrum but exhibit larger widths where shock front/cloud interactions contribute to high-velocity [O IV] components.

\begin{figure}
	\figurenum{10}
	\centering
		\includegraphics[width=0.4\textwidth,angle=90]{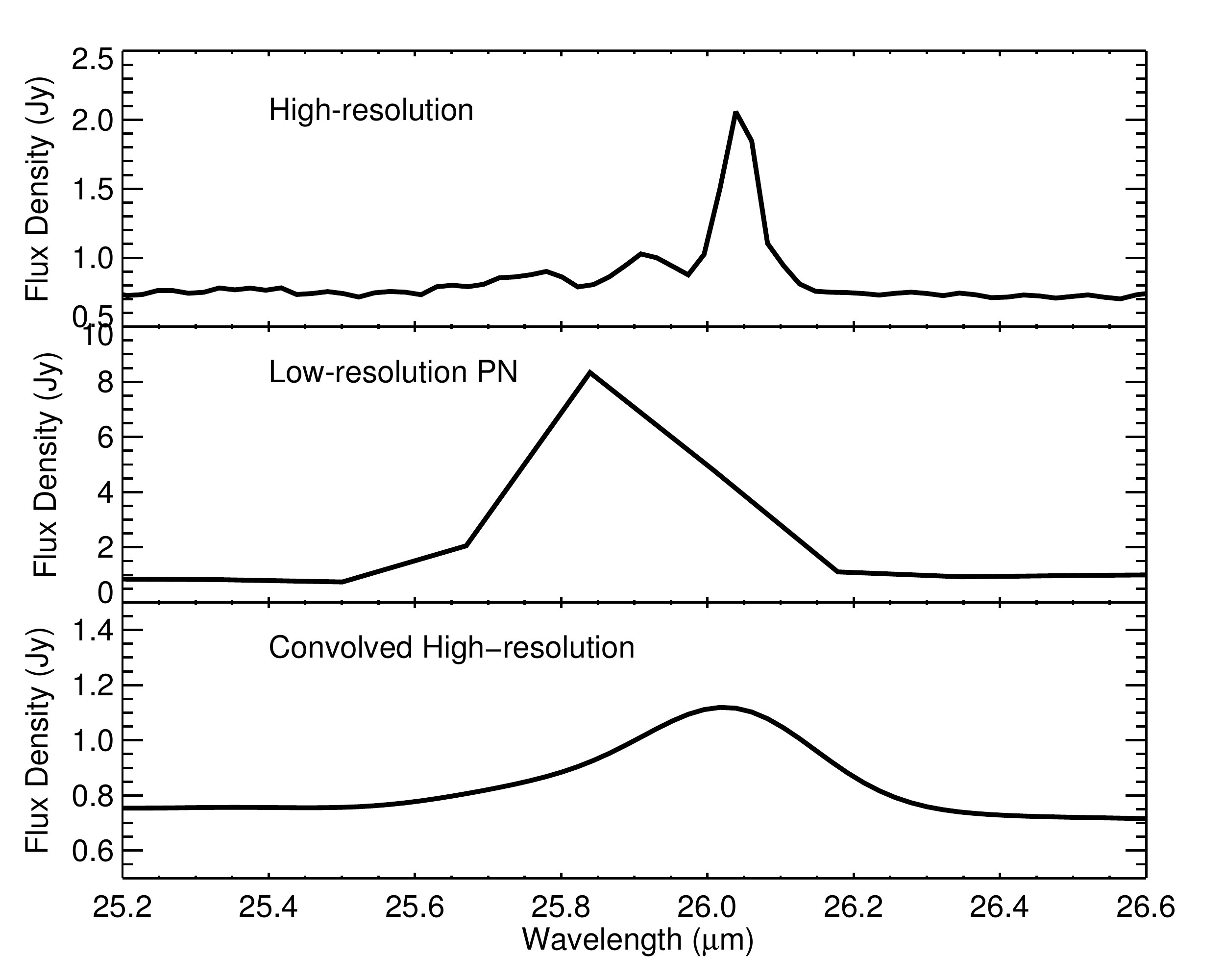}
	\caption{\scriptsize (Top) Sample spectrum from scan 2 of high-resolution [O IV] (Middle) Sample low-resolution spectra taken from a \textit{Spitzer} IRS image of PNG 002.7-04.8 (Bottom) The top two spectra convolved.}
\end{figure}

\begin{deluxetable} {ccccc}
\tabletypesize{\footnotesize}
\tablecaption{Gaussian fit values for convolved high-resolution [O IV] spectra}
\tablewidth{0pt}
\tablehead{
Position & C$_{0}$ & C$_{2}$ & Integrated width & Velocity Width\\
& (Jy) & ($\mu$m) & ($\mu$m) & (km s$^{-1}$)
}
\startdata
shock1-pos01 & 0.19 & 0.20 & 0.21 & 2400\\
shock1-pos02 & 0.14 & 0.24 & 0.25 & 2900\\
shock2-pos01 & 0.37 & 0.17 & 0.18 & 2100\\
shock2-pos02 & 0.25 & 0.18 & 0.19 & 2200\\
shock2-pos03 & 0.15 & 0.19 & 0.21 & 2400\\
shock2-pos04 & 0.085 & 0.24 & 0.26 & 3000\\
\enddata
\end{deluxetable}

From the convolved high-resolution spectrum, we evaluated the width of the [O IV] features using a five-parameter Gaussian,

\begin{equation}
\mbox{flux} = C_{0} \: \exp(-0.5 z^{2}) + C_{3} \lambda + C_{4}, \: z = \frac{\lambda - C_{1}}{C_{2}}
\end{equation}

\begin{figure*}
	\figurenum{11}
	\centering
		\includegraphics[width=0.57\textwidth,angle=270]{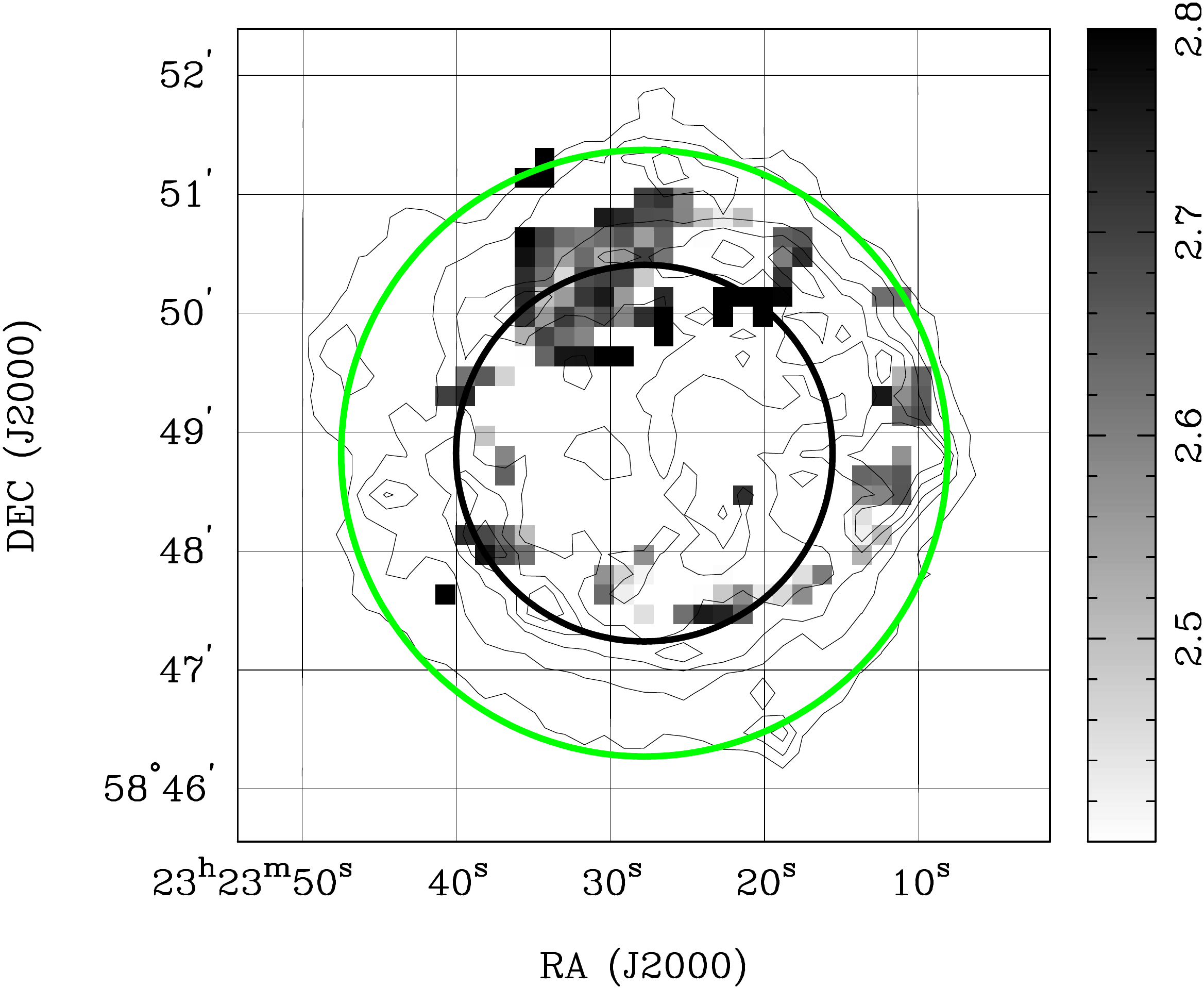}
	\caption{\scriptsize Low-resolution [O IV] (25.9 $\mu$m) \citep{smith09} line widths given in $10^{3}$ km s$^{-1}$ with VLA 20 cm contours overlaid.  The image has been masked for regions where the [O IV] integrated intensity was at least $3-\sigma$ the noise level, or about 10\% of the maximum value of  $8 \times 10^{-6}$ W m$^{-2}$ sr$^{-1}$.  Here we show regions with the largest [O IV] line widths, exceeding values inferred from our convolved high-resolution spectra ($\sim 2400$ km s$^{-1}$).  The black circle represents the approximate location of the reverse shock at $95''$ from the expansion center \citep{thor01,del10} while the green circle represents the forward shock.  Note that in some places, the broad [O IV] emission extends well beyond the reverse shock, thus it is unlikely to be associated with the Bright Ring.}
\end{figure*}

\noindent This profile proved insufficient as there was significant broadening in the wings of the [O IV] feature and some of the lines appeared distinctly non-Gaussian.  Instead, we used this fit to subtract a linear baseline from the data and integrated the intensity over the wavelength range of the line profile.  We multiplied this value by $\frac{1.17441}{\pi C_{0}}$, which for a purely Gaussian function would yield the full-width at half-maximum.  In our data set, this value is comparable to the line width although less biased by non-Gaussian line profiles.  The values of $C_{0}$, $C_{2}$, the calculated line width, and the reduced chi-square of the Gaussian fit are provided in Table 2 for each pointing in our high-resolution data where we produced a convolved low-resolution fit. We infer from this analysis that the average line width in the region of our high-resolution spectroscopy is on the order of $2400$ km s$^{-1}$.

From the low-resolution [O IV] spectra of Cas A, we constructed a data cube of the remnant using IRS CUBISM \citep{smith07}.  The line width of [O IV] around the remnant can then be computed from the second moment of the spectrum at each position.  In many cases, the signal to noise of the [O IV] line was too low to compute a reliable line width, so we masked every position in our data cube whose integrated [O IV] intensity was less than the $3-\sigma$ the level of the noise, or less than 10\% of the maximum value of $8 \times 10^{-6}$ W m$^{-2}$ sr$^{-1}$.  In Figure 11, we show the [O IV] line width (in $10^{3}$ km s$^{-1}$) in regions where it exceeds our average line width of $2400$ km s$^{-1}$.

There is some correlation between positions of broadening in [O IV] (Figure 11) and CO broadening (Figure 4) as well as regions of bright radio emission.  Highly broadened [O IV] emission to the north, west, and southeast of Cas A support the hypothesis that ejecta in these regions are interacting with interstellar gas.  The huge velocity spread of the [O IV] line presumably arises at the outward edge of the shock where gas is being stirred strongly by the interaction, whereas the CO lines must correspond to greater depths in the molecular cloud where the effects of the shock are just being felt.  In some regions, broad [O IV] line profiles correspond to the location of the Bright Ring, where a large velocity spread can be attributed to the large quantity of ejecta brightened by the reverse shock.  We can distinguish these two regions by their position relative to the positions of the reverse shock (Figure 11, black circle) and forward shock (green circle).  Some of the broadened ejecta to the north, west, and southeast of Cas A correspond to positions well beyond the reverse shock, and in some cases, at projected radii consistent with the forward shock.  We infer that the broad [O IV] line ejecta at these positions are associated with turbulence due to interaction with surrounding gas.

\section{Discussion}

Our study indicates that any signs of interaction between the shock from Cas A and the surrounding ISM are subtle.  Broadened CO lines extending beyond the shock front offer the most direct evidence in support of an interaction between high-velocity ejecta (i.e. ejecta with velocities exceeding the spherical expansion velocity of Cas A) and molecular gas. We discuss this behavior first, in Section 5.1. In Section 5.2 we discuss the mid-infrared lines and the evidence for turbulence they provide. We describe aspects of the SNR/ISM interaction in other supernovae and compare the physics of these interactions to Cas A in Section 5.3.

\subsection{Possible Mechanisms for CO Line Broadening in Regions Beyond the Cas A Shock Front}
 
Galactic molecular clouds typically have linewidths for $^{12}$CO of $1 - 3$ km s$^{-1}$ \citep{hb04}.  We have observed linewidths as large as $8 - 10$ km s$^{-1}$ in regions that extend $1 - 2'$ past the southwest shock front of Cas A.  We infer that, given the small projected distance between the gas and Cas A and LSR velocities of features in these clouds similar to those observed in previous studies of gas around Cas A \citep{ll99}, there must be some mechanism associated with the SNR other than direct interaction with the shock front that accounts for these linewidths.  In the following discussion, we consider two possible causes of this behavior.  Two remaining possibilities that we consider less likely are discussed in Appendix B.  We conclude that the most likely cause of line broadening in the observed molecular clouds is interaction between the clouds and knots of material ejected at higher velocities than that of the shock front.

\subsubsection{Broadening by Stellar Winds}

It has been proposed that the Cas A progenitor was a star on the order of $15 - 40 \mbox{ M}_{\odot}$ \citep{young06}.  For stars in this range, episodes of significant stellar winds can be described by the prescriptions of OB mass loss, red giant/supergiant mass loss, and Wolf-Rayet (WR) mass loss leading to a total mass loss of $5 - 10 \mbox{ M}_{\odot}$.  We consider a molecular cloud that intercepts $\sim 10\%$ of the momentum from the wind (i.e. for wind emitted isotropically, a cloud with radius of $1.9$ pc at $3$ pc distance from the supernova progenitor would intercept approximately 10$\%$ of the wind).  We then estimate that this wind would need to emerge with a velocity in excess of $2000$ km s$^{-1}$ to provide the momenta we calculated in each of our molecular clouds.  This wind velocity is at the upper limit for the terminal velocity of winds emitted via OB mass loss \citep{kp00} and WR mass loss \citep{crow07}, let alone red giant/supergiant mass loss which involves significantly slower terminal velocities.  

Only in the most extreme mass loss scenario does the assumed velocity outflow match theoretical models.  For example, \citet{young06} predict that for a $40 \mbox{ M}_{\odot}$ progenitor, mass loss in the form of Luminous Blue Variable (LBV) eruptions and subsequent WR mass loss would leave a pre-supernova collapse mass of $7.8 \mbox{ M}_{\odot}$.  Ejecta mass on the order of $30 \mbox{ M}_{\odot}$ combined with a wind velocity on the order of $2000$ km s$^{-1}$ for LBV winds approximately matches the momentum in turbulent molecular line widths observed in our molecular clouds, however the extreme mass-loss in this progenitor might uncover its C/O core before collapse.  Observation of nitrogen-rich knots in the ejecta of Cas A preclude such a massive progenitor \citep{fes01b}.  For example, recent models obtained from diffuse, thermal X-ray emission suggest that Cas A mass loss occurred via a slow RSG wind ($\sim 15$ km s$^{-1}$) with only a moderate total mass stripped off the progenitor star ($\sim 11 \mbox{ M}_{\odot}$) and even less in the wind itself ($\sim 6 \mbox { M}_{\odot}$) \citep{lphs13}.  Furthermore, while there is evidence that the Cas A progenitor exploded into a circumstellar bubble created by stellar winds \citep{rhfw95, bsbs96}, models developed from observations of the interaction between the remnant and this swept up material do not provide enough velocity in the wind \citep{pgl09} to account for the CO line widths observed.  We infer that while there is evidence for a circumstellar wind being swept up by Cas A, the wind could not account for the anomalous line profiles observed in molecular clouds toward the SNR.

\subsubsection{Broadening by Interaction with Outflows of Fast-Moving Knots}

There is evidence in optical and X-ray observations of Cas A that some fast-moving knots are moving along bipolar outflows toward the southwest and northeast of the SNR \citep{fes01, fes06, del10}.  In addition, there are ``ejecta pistons'' \citep{del10} of fast-moving knots that extend in other directions besides along the outflows.  These ejecta are characterized especially by iron, silicon, and oxygen emission at velocities on the order of $4000 - 10000$ km s$^{-1}$, appreciably in excess of the Cas A forward shock velocity.  The bright optical and X-ray emission extends in many places $1 - 2'$ in projected radius beyond the forward shock.  Along the observed outflow, points of bright emission form a cone with a projected opening angle calculated at $27^{\circ}$ \citep{del10}.

Our observations are consistent with shock-broadened molecular lines caused by the outflowing fast-moving knots.  These knots have been observed at radii consistent with the projected radius of the observed molecular clouds.  The energy in the fast-moving knots could be sufficient to create the broad lines.  Less than one percent of the $2 \times 10^{51}$ ergs in the Cas A supernova event \citep{young06} would be sufficient to account for the $\sim 10^{49}$ ergs (i.e. a $10^{4} \mbox{ M}_{\odot}$ cloud with $10$ km s$^{-1}$ linewidths) of turbulent energy in the molecular cloud.  For example, \citet{del10} calculate that the fast-moving outflow carries $3 \times 10^{50}$ ergs in kinetic energy, roughly $3\%$ of which needs to be intercepted by one of the observed clouds to account for the turbulent energy we infer.  Assuming the outflow is projected along a cone with opening angle $27^{\circ}$ out to a distance of $8$ pc (i.e. 2 pc beyond the shock front of Cas A), a cloud with a radius of $1.2$ pc would intercept roughly $9\%$ of the incident ejecta.

\subsection{Turbulence in Mid-Infrared Line Emission}

We have found that emission lines observed toward the Cassiopeia A shock front exhibit multiple, distinct Doppler components.  Multiple optical and infrared studies of Cas A have revealed knots of ejecta at velocities ranging from $-2300$ to $3200$ km s$^{-1}$ \citep[our Figure 7;][]{law95, rhfw95, del10}.  These knots reveal a complex, three-dimensional picture of the remnant, although one in which most of the expanding ejecta exhibit roughly spherical geometry with broad line regions located toward the optical shell.  This characteristic of Cas A spectra supports the interpretation of extreme turbulence in bright, high-velocity [O IV] emission along the northern shock front.  Here we will explore the possible mechanisms that may account for the conversion of the kinetic energy into turbulence.

The physical scenario represented by turbulent line profiles in the forward shock of a SNR is at odds with the canonical model of a freely-expanding remnant.  Any initial turbulence in the ejecta provided by the supernova event will rapidly dissipate unless some additional interaction can provide energy to the ejecta in order to excite turbulent motion on smaller scales.  The most plausible candidates for this interaction are MHD driven turbulence due to the magnetic field inherent in the shock itself or a direct interaction between the forward shock and dense gas in the interstellar medium.

To evaluate the first possibility, the magnetic field in the Cas A shock front can be indirectly probed through X-ray synchrotron emission driven by first-order Fermi acceleration \citep{vl03}.  Turbulence generated by, for example, the interaction between diffusive shock driven cosmic rays and the supernova ejecta may account for the observed magnetic field `frozen-in' to plasma in the ejecta \citep{vm81, bl01, bell04}.  In this way, magnetic fields provide a direct probe of the energy necessary to drive turbulence in the shock front.  The observed magnetic field toward the Cas A shock front has a strength measured between 0.08 mG \citep{vl03} and 0.5 mG \citep{saha14}.  We approximate the average turbulent line velocity as,

\begin{equation}
v_{\mbox{\tiny turb}} = \sqrt{\frac{B^{2}}{\mu m_{\mbox{\tiny H}} n_{\mbox{\tiny eff}}}}
\end{equation}

If we take 0.5 mG as a rough upper limit for the magnetic field strength at the shock front and assume a gas density on the order of $n_{\mbox{\tiny eff}} = 30$ cm$^{-3}$ \citep{saha14}, we can calculate an energy density in the magnetic field and assume this value is comparable to the energy density in turbulence.

The calculated turbulent velocities are no more than $200$ km s$^{-1}$ which is significantly slower than the observed turbulence.  We infer that any turbulence related to interactions with the `frozen-in' magnetic field in the ejecta plasma is insufficient to explain the observed high-velocity features.

The second mechanism that may generate turbulence in supernova ejecta is interaction with large scale density fluctuations in the medium surrounding Cas A.  As the Cas A shock front freely expands outwards, interaction between the ejecta and interstellar material will provide turbulent energy such that high-velocity Doppler components and large line widths persist in the fine structure lines toward the shock front.

A direct interaction between the SNR and surrounding gas explains both the sharp cut off in the spectral lines and the lack of change in the relative strength of the high-velocity features at different positions.  Young SNRs exhibit a contact discontinuity in their radial density profile between ejecta and circumstellar or interstellar gas behind which turbulence associated with Rayleigh-Taylor instabilities generate magnetic fields \citep{gull75}.  The dominant mechanism for radio emission in Cas A is believed to be synchrotron radiation generated by an interaction between relativistic electrons in the ejecta and the magnetic field behind this contact discontinuity \citep{rosen70, kgp98}.  Thus, a sharp contact discontinuity between Cas A and the interstellar gas would exhibit the cut off in line strength we observe over a relatively small ($\sim 10''$) transverse distance.  Furthermore, the consistently high turbulent velocities over this distance indicate that the interaction between Cas A and the surrounding gas has stirred all the observed layers of ejecta recently such that the power in turbulent motion in the ejecta has not had time to dissipate from this position in the shock front.

We can point to other signatures as an indication of interaction between ejecta and swept up interstellar or circumstellar material.  The bright radio emission behind the northern, western, and southern shock front of Cas A suggests that either stronger magnetic fields or more efficient particle acceleration in this region must be invoked to explain the strong synchrotron component \citep{del02}.  We infer that an interaction between the ejecta and swept up material both causes the turbulent line velocities and forms a shock interface where efficient Fermi acceleration occurs.  This explanation would also agree with a scenario where the forward shock is beginning to form an interface with a molecular cloud located toward this region of the SNR.  Other regions where there exist both broadened molecular lines and bright radio emission could be sites of a contact discontinuity between the Cas A forward shock and molecular gas in the interstellar medium.

There does not appear to be significant CO line broadening toward the cloud observed north of Cas A, despite the broad range of velocities in the IR fine-structure lines. It appears that the shock has not reached the bulk of the mass of the molecular cloud, represented by the CO. To understand the situation, we estimate the relative masses involved in the two types of emission. \citet{hines04} found that the CO-emitting molecular cloud is about $650$ M$_{\odot}$. In the appendix, we show that the column density of ejecta along the northern shock front is $N = 1.1 \times 10^{20}$ cm$^{-2}$. Assuming a distance of $3.4$ kpc, we find that approximately $0.05$ M$_{\odot}$ of material within the IRS slit would be sufficient to produce the infrared lines at position S2, P1. That is, allowing for ten IRS slit areas of emission (about 10 by 100 arcsec), some three orders of magnitude less material is required for the infrared lines as is needed to produce the CO emission. Therefore, the observations are consistent either with the infrared lines arising through interaction with a much smaller body of gas isolated from the main CO-detected cloud, or with a situation where the shock has just reached the cloud and has only penetrated slightly into it. The latter possibility would be consistent with Figures 1 and 3, which show the enhanced radio synchrotron emission indicative of the interaction region only lying along one edge of the cloud.

\subsection{Comparison to Other Studies of SNR/MC Interactions}

Previous studies of SNR/MC interactions focus on several signatures of shocked molecular gas.  The presence of the broadened molecular lines, the OH 1720 MHz maser, and near-IR molecular hydrogen lines are common tracers that distinguish regions of shocked gas around an expanding SNR.  We describe the well-observed SNRs IC443, W44, and W28 and the tracers used to identify interaction with molecular clouds in their vicinity for comparison with Cas A.  

While these objects are at approximately the same distance as Cas A they are considerably older.  IC 443 is 1.5 kpc and 4000 yr \citep{raa07, loz81, troj08}, W44 is 3 kpc and 20,000 yr \citep{wcd91}, and W28 is 2.8 kpc and 58,000 yr \citep{kas93}.  Age is an important factor in evaluating the characteristics of SNR/MC interactions for two reasons.  First, Cas A is smaller than the more evolved examples of SNR/MC interactions.  The entire SNR measures approximately $5'$ in diameter or about $5$ pc at $3.4$ kpc distance.  Shocks caused by the blast wave from Cas A are less likely to have propagated through the molecular cloud, so any observed variations in the turbulent linewidths in these clouds are more likely to be confined to regions close to the Cas A shock front.  Second, Cas A is still expanding at approximately the free expansion velocity ($3000 - 4000$ km s$^{-1}$).  For hydrodynamic models with shock speeds $> 25$ km s$^{-1}$, molecular clouds exhibit significant collisional dissociation of H$_{2}$ molecules \citep{kwan77}.  Therefore, any interaction with molecular gas is likely to dissociate gas at the point of interaction between the two media.   As the shock wave from an interaction propagates into a molecular cloud, emission lines from any surviving molecules will be broadened relative to the un-shocked material.  This phenomenon is observed for a number of other ``broad molecular line'' sources, including 3C391 \citep{rr99} and HB21 \citep{koo01}.

The SNR/MC interaction for IC 443 is thoroughly studied and is useful to compare with the younger interaction near Cas A. The interaction for IC 443 was first identified by \citet{cornett77} and has been confirmed by detection of molecular species toward the associated molecular clouds \citep{white86, xwm11}, shocked OH, CO, and H$_{2}$ emission \citep{den79a, den79b, burt88}, and shock excitation of the 1720 MHz OH maser \citep{claus97, hew06}.  The physical conditions in this SNR appear to be well-suited to detect tracers of SNR/MC interactions.  The interacting molecular cloud is in front of the remnant along the line of sight and possesses more mass ($> 10^{4} \mbox{M}_{\odot}$) than the clouds detected near Cas A.  The cloud exhibits linewidths of $> 30$ km s$^{-1}$, up to $90$ km s$^{-1}$ in regions with the most turbulent molecular gas \citep{white86}.  This result has been confirmed in several molecules, probing molecular gas of densities from $10^{5}$ cm$^{-3}$ to $3 \times 10^{6}$ cm$^{-3}$ \citep{vjp93}.  The same result has been observed in W44 and W28, where CS, CO, and HCO$^{+}$ line widths are $20 - 30$ km s$^{-1}$ \citep{rrj05}.  The resulting turbulence driven fragmentation may suppress star formation in regions of giant molecular clouds where SNR/MC interactions occur; smaller clouds have also been identified around IC 443, down to mass scales of $0.1 - 0.3 \mbox{ M}_{\odot}$.

Shock excitation in the SNR/MC interactions located toward IC 443 cause this remnant to be one of the most luminous molecular hydrogen sources in the galaxy \citep{burt88}.  Since it is difficult to excite bright emission in this species without dissociating the molecule, near-IR molecular hydrogen emission is predominantly useful in more evolved SNR as a tracer for SNR/MC interactions.  In IC 443, the correlation between line emission from H$_{2}$ $v = 1-0 \mbox{ S}(1)$ and the more easily detected CO, HCO$^{+}$, and HCN emission at longer wavelengths provides a basis for mapping the region where the SNR is interacting with dense molecular gas.

The OH 1720 MHz maser is one of the most commonly used tools to identify SNR/MC interactions.  Spectral line emission from this phenomenon is suited to more evolved SNRs where C-type (nondissociative) shock interfaces are observed between the supernova and a molecular cloud.  Masers around W28 and W44 are observed to have sizes of 150 - 1000 AU \citep{hgbc05} with cores 60 AU in size.  In these regions where the number density of molecular hydrogen is on the order of $n_{H_{2}} \sim 10^{4}$ cm$^{-3}$, the shock velocity cannot exceed $10$ km s$^{-1}$.  The physical conditions that give rise to the OH $1720$ MHz maser are so specialized that they occur only in a small fraction of the remnant volume and these conditions may not appear at all in small and young remnants.

Cas A is a very young SNR ($\sim 350$ yr) with a high shock velocity ($3000 - 4000$ km s$^{-1}$).  Any interaction between the SNR and nearby molecular clouds will be at an early stage of development considering that the diameter of Cas A is comparable to the size of a typical $10^{4} \mbox{ M}_{\odot}$ molecular cloud.  Our observations do show subtle indications of an interaction between the SNR and molecular clouds.  These indicators include the extended region of broadened CO lines, and the multiple high-velocity components of the mid-infrared fine structure lines around the rim of the SNR.  However, the lack of more dramatic tracers, such as even broader CO lines, bright emission from molecular hydrogen, and OH maser emission can all plausibly be explained by the youth and small size of the Cas A remnant or, alternatively, a small amount of mass in the fastest ejecta knots.

\section{Conclusion}

From our OTF maps in $^{12}$CO and $^{13}$CO J $=2-1$ of the region around Cas A, and high-resolution spectroscopy from $10-40$ $ \mu$m across two tracks in declination across the northern shock front of Cas A, supplemented by a VLA 20 cm map of Cas A \citep{arlpb91} and low-resolution spectroscopy of [O IV] (25.94 $\mu$m) \citep{smith09}, we find the following:

\begin{enumerate}
\item Bright radio continuum emission along the northern edge of Cas A toward a previously identified molecular cloud suggests acceleration of particles at the shock front/cloud interface.

\item There is broadening of $^{12}$CO J$=2-1$ in CO spectral features across the western and southern shock fronts of Cas A.  The presence of broadened molecular gas that extends well beyond the boundaries of the SNR in these directions suggests an interaction between Cas A and the molecular gas.  We suggest that a shock interaction with ejecta in the fast-moving knots, which emerge at velocities substantially greater than that of the shock front, has caused the observed CO line broadening.

\item There are a number of mid-infrared fine-structure lines across the northern shock front of Cas A.  Some of these lines exhibit significant high-velocity splitting that is not accounted for by typical supernova outflows.  These lines suggest an interaction with interstellar material, with the resulting turbulence yielding a distribution of large radial velocities.

\item Broadening in the low-resolution spectra of [O IV] confirms that high velocities occur along the northern shock front of Cas A as well as along the western and southern shock fronts in the same regions where we observe CO broadening.
\end{enumerate}

There is subtle evidence in our observations for an interaction between Cas A and the interstellar medium along the northern shock front as well as toward molecular clouds located to the west, south, and southeast.  Our interpretation is consistent with previous submillimeter and mid-infrared studies of Cas A as well as optical and X-ray observations of fast-moving knots toward the remnant.

\acknowledgments

The Heinrich Hertz Submillimeter Telescope is operated by the Arizona Radio Observatory with partial support from National Science Foundation grant AST 1140030.  Funding for this research was provided by NASA through Contract Number 1255094 issued by JPL/Caltech.  We would like to express our gratitude to Nathan Smith for his helpful insight on the underlying physics of our detections as well as Oliver Krause for his input.

\begin{figure}
	\figurenum{12}
	\centering
		\includegraphics[width=0.42\textwidth]{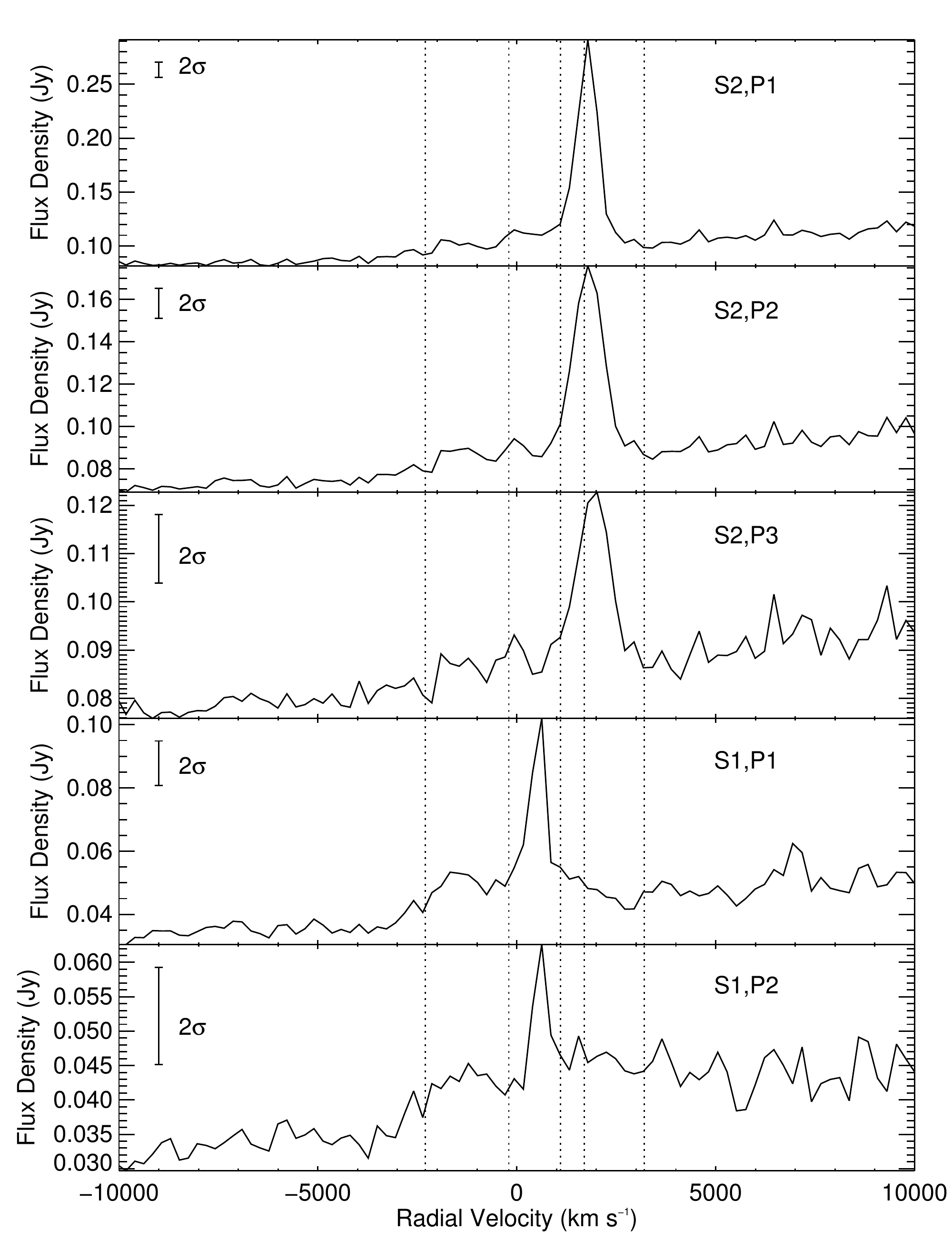}
	\caption{\scriptsize Identified [S III] ($18.71$ $\mu$m) lines in scans 1 and 2 (S1 and S2) at the southernmost pointings in each shock (P1-2 in S1, P1-3 in S2).  These lines have been plotted versus relative velocity to demonstrate the presence of high-velocity components in the spectra.  A sample error bar is shown in the upper-left hand corner to compare with the strength of the identified features.  Five dotted, vertical lines mark some recurrent high-velocity features (at $-2300, -200, 1100, 1700, 3200$ km s$^{-1}$).}
\end{figure}

\appendix

\section{Mid-Infrared Fine Structure Lines}

We discuss the velocity structure and other aspects of mid-infrared fine structure lines other than [O IV] in this appendix.  These lines have been observed previously in Cas A \citep{ennis06, rho08}.  In this appendix, we present the deepest mid-infrared spectroscopic analysis performed on this remnant.

\subsection{[S III] 18.71 \texorpdfstring{$\mu$}{}m line}

The second brightest fine-structure line after [O IV] is the [S III] feature at $18.71$ $\mu$m.  In scan 1, there are three unique spectral lines around the [S III] rest wavelength centered at $18.65, 18.77, 18.81$ $\mu$m (Figure 12).  These wavelengths correspond to radial velocities of $-1600$, $-960$, and $960$ km s$^{-1}$.  These values do not appear to correspond closely with any of the velocities calculated for the [O IV] emission, however, the limits to our spectral resolution and the poor detection of [S III] in these frames may account for the discrepancy.

In scan 2, [S III] was more strongly detected in each frame and across more pointings.  The high-velocity components in [S III] correspond to three distinct spectral features centered around the rest wavelength.  These features are centered at $18.78, 18.89, 19.03$ $\mu$m corresponding to radial velocities of $-5130, -2880, -1120$ km s$^{-1}$.  This pointing yields the largest absolute radial velocities detected in any line, although the two smaller velocities agree with measurements of the [O IV] line.

\subsection{[S IV] 10.51 \texorpdfstring{$\mu$}{}m line}

Fine structure emission was observed around $10.5$ $\mu$m along both scans, indicating some emission due to [S IV].  At scan 1, we observed a single Doppler component around [S IV] (Figure 13) where the line is dominated by the brightest component at $10.54$ $\mu$m, corresponding to radial velocity of $960$ km s$^{-1}$.

\begin{figure}
	\figurenum{13}
	\centering
		\includegraphics[width=0.42\textwidth]{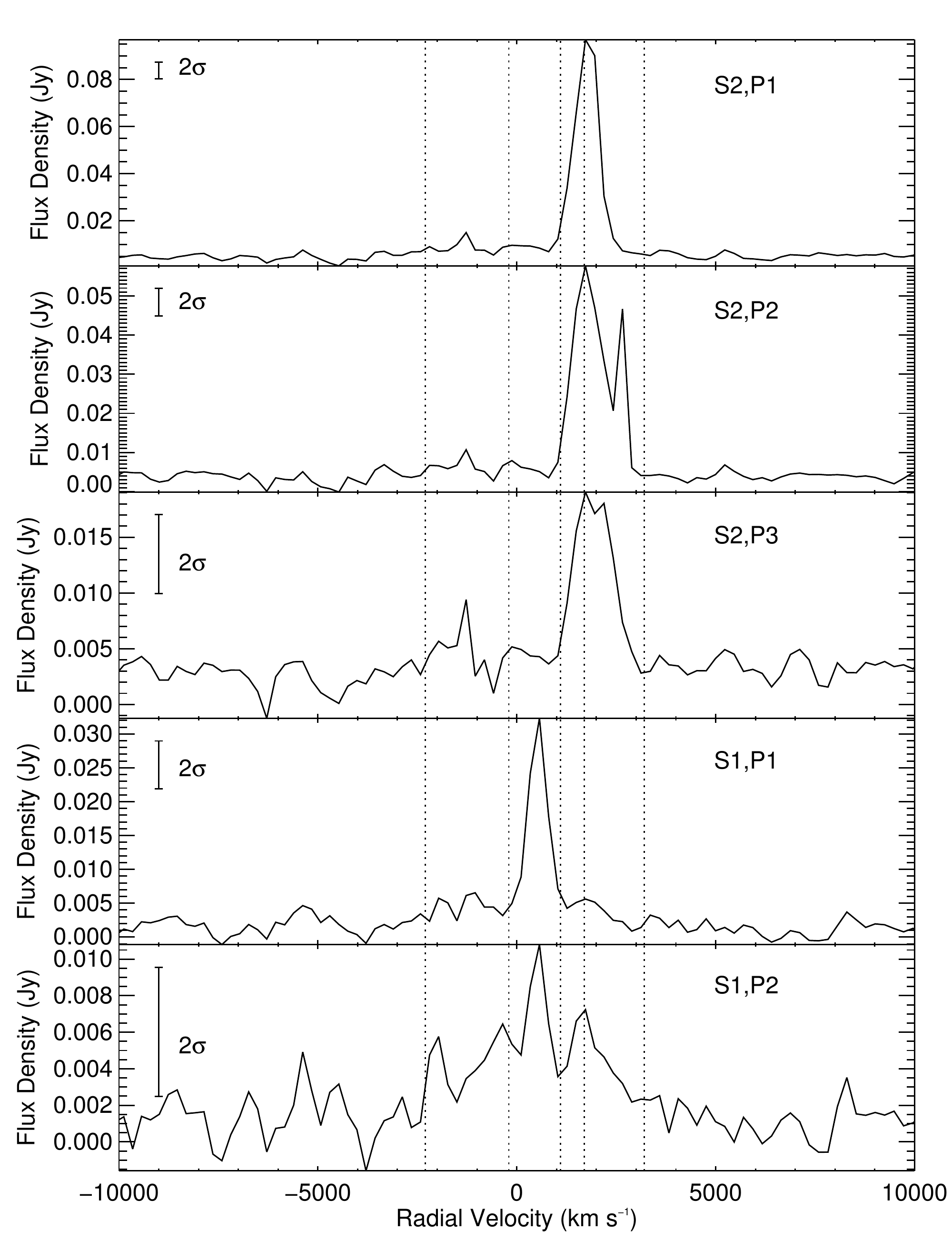}
	\caption{\scriptsize Identified [S IV] ($10.51 \mu$m) lines in scans 1 and 2 (S1 and S2) at the southernmost pointings in each shock (P1-2 in S1, P1-3 in S2).  These lines have been plotted versus relative velocity to demonstrate the presence of high-velocity components in the spectra.  A sample error bar is shown in the upper-left hand corner to compare with the strength of the identified features.  Five dotted, vertical lines mark some recurrent high-velocity features (at $-2300, -200, 1100, 1700, 3200$ km s$^{-1}$).}	
\end{figure}

Similarly, in scan 2 there is very little high-velocity splitting, and stacking the spectra reveals only one main component due to [S IV] at $10.57 \mu$m corresponding to a radial velocity of $1700$ km s$^{-1}$.  This result is anomalous considering that we detected a component of differing velocity at pointing 1 and the critical densities of [S III] and [S IV] are essentially the same yet we have found only negative velocities for [S III].  We infer that there may be some shock-excitation of high-velocity features apart from the dominant components observed in the [S IV] line. 

\subsection{[Ne II] 12.81 \texorpdfstring{$\mu$}{}m line}

Figure 14 shows that there are three distinct high-velocity features at both scans.  In scan 1, we identified two blue-shifted features centered at $12.74$ and $12.76$ $\mu$m and a third feature at $12.80$ $\mu$m.  These wavelengths correspond to radial velocities of $320$, $1260$, and $1730$ km s$^{-1}$.  However, this identification is anomalous in that the blue-shifted features have approximately the same flux as the low-velocity feature and do not correspond to any feature observed in [S III] or [S IV].

\begin{figure}
	\figurenum{14}
	\centering
		\includegraphics[width=0.42\textwidth]{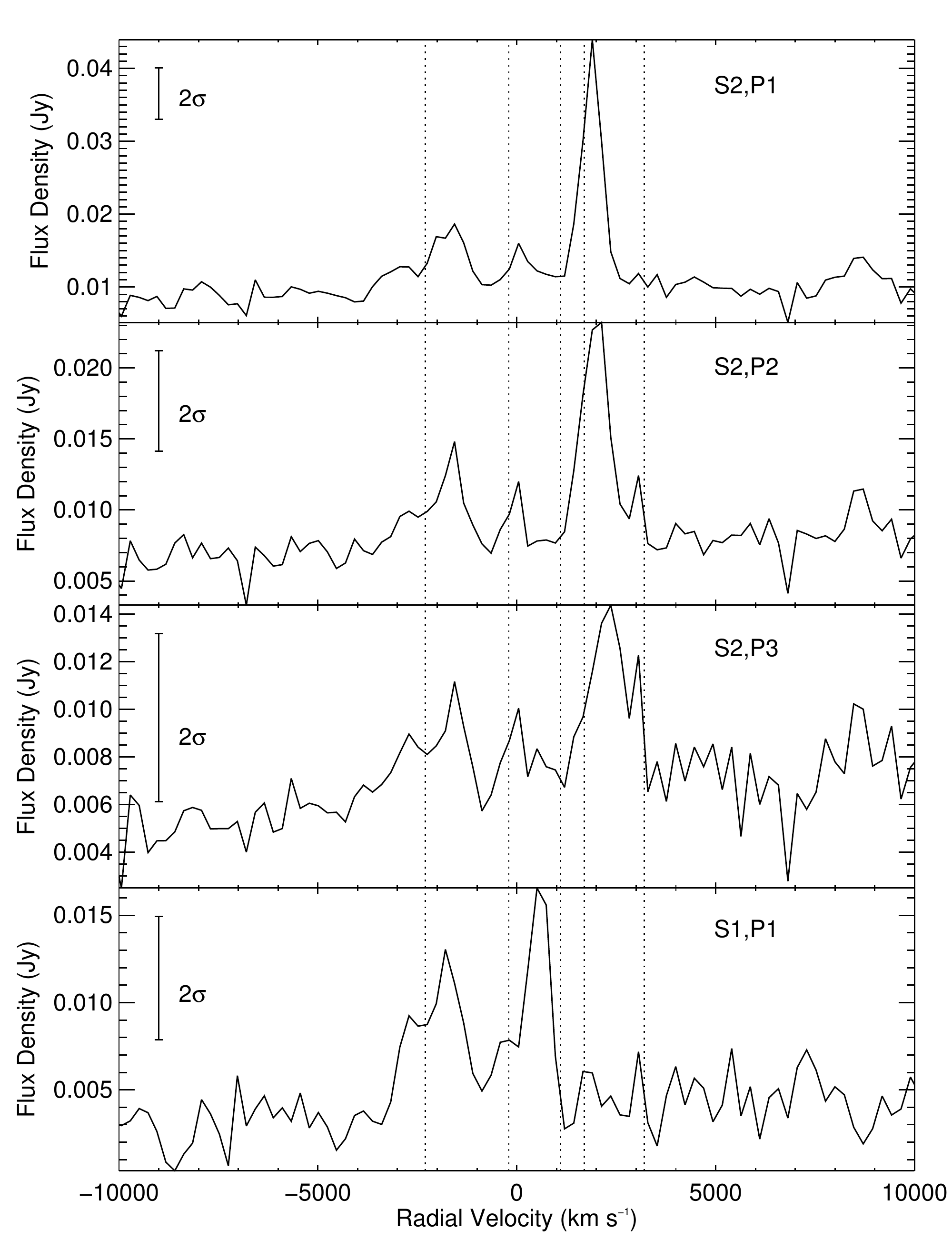}
	\caption{\scriptsize Identified [Ne II] (12.81 $\mu$m) lines in scans 1 and 2 (S1 and S2) at the southernmost pointings in each shock (P1 in S1, P1-3 in S2).  These lines have been plotted versus relative velocity to demonstrate the presence of high-velocity components in the spectra.  A sample error bar is shown in the upper-left hand corner to compare with the strength of the identified features.  Five dotted, vertical lines mark some recurrent high-velocity features (at $-2300, -200, 1100, 1700, 3200$ km s$^{-1}$).}	
\end{figure}

There is some agreement with the velocities observed in scan 1 for [O IV] and [S III]; assuming an error of roughly $200$ km s$^{-1}$ per measurement (given the spectral resolution of the IRS high-resolution modules), the [O IV] feature at $1740$ km s$^{-1}$ and [Ne II] feature at $1730$ km s$^{-1}$ likely correspond to the same high-velocity component, as well as the [O IV] feature at $340$ km s$^{-1}$ and the [Ne II] feature at $320$ km s$^{-1}$.  Furthermore, the [S III] feature at $960$ km s$^{-1}$ and [Ne II] feature at $1260$ km s$^{-1}$ may correspond to the same component.

In scan 2, the high-velocity structure exhibits a similar pattern with a bright peak at low-velocity and two distinct features that are both blue-shifted.  These features occur at wavelengths $12.74$ and $12.78$ $\mu$m for the blue-shifted peaks and $12.82$ $\mu$m for the low-velocity peak.  These wavelengths correspond to radial velocities of $1730$, $800$, and $-140$ km s$^{-1}$.  We can associate the [O IV] $1510$ km s$^{-1}$ feature with [Ne II] at $1730$ km s$^{-1}$ within a reasonable margin of error in scan 2, as well as the [S IV] $-290$ km s$^{-1}$ and [Ne II] $-140$ km s$^{-1}$ features.

It is puzzling that [Ne II] would exhibit exclusively blue-shifted features at high-velocity where the other lines indicate that there are red-shifted shock components.  Furthermore, [Ne II] exhibited multiple blue-shifted components in scan 2 where [S IV] and [S III] have none and [O IV] appears at only the highest velocity feature.  These observed spectral properties must be explained by any model for the shocked gas.

\subsection{Remaining emission lines}

The remaining two emission lines [Ne V] at 14.32 $\mu$m (Figure 15) and [Ne III] at 15.56 $\mu$m (Figure 16) do not show any high-velocity structure above the noise level.  Both lines have distinct peaks between $2000$ and $3000$ km s$^{-1}$.

\begin{figure}
	\figurenum{15}
	\centering
		\includegraphics[width=0.42\textwidth]{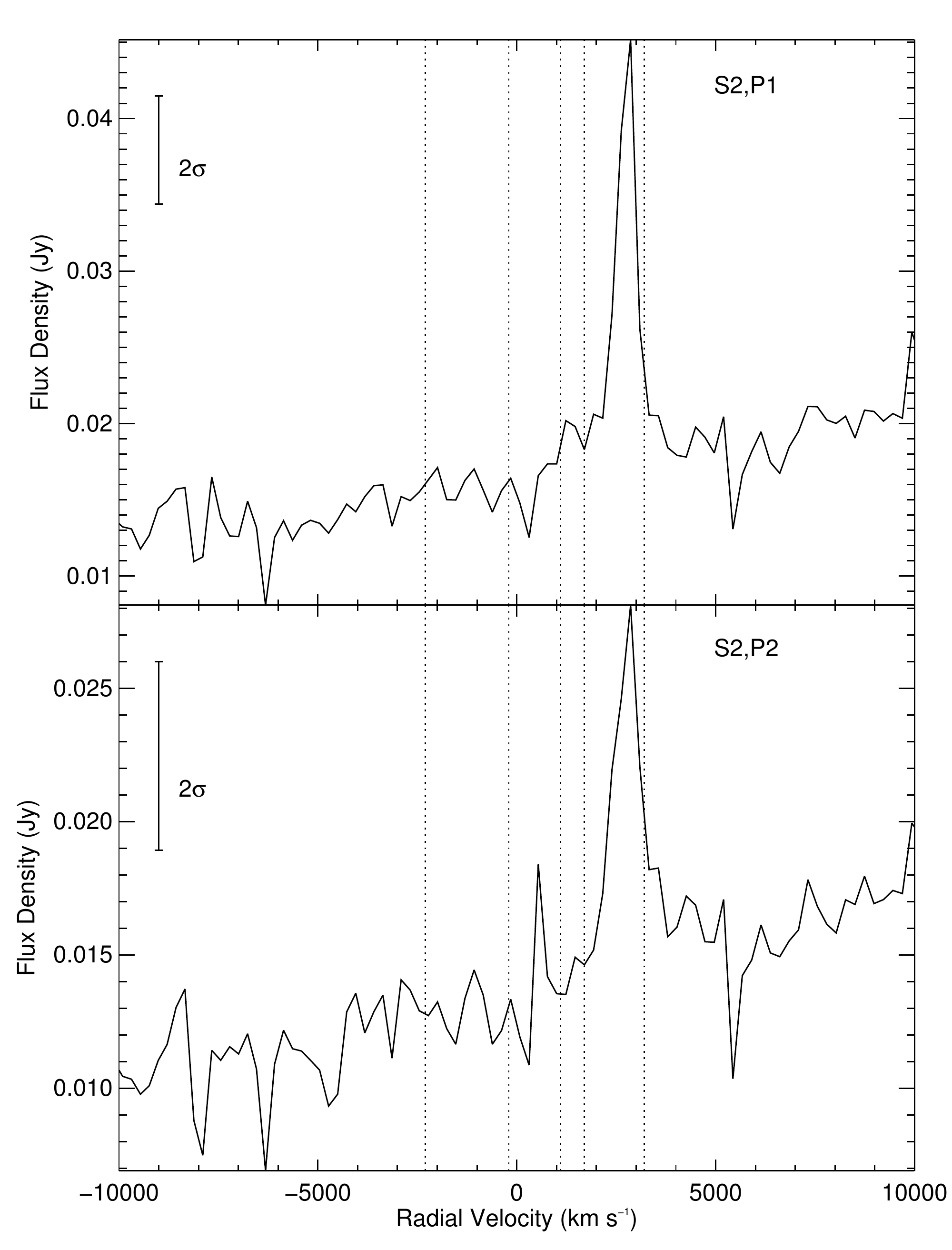}
	\caption{\scriptsize Identified [Ne V] (14.32 $\mu$m) lines in scan 2 (S2) at the southernmost pointings (P1-2).  These lines have been plotted versus relative velocity to demonstrate the presence of high-velocity components in the spectra.  A sample error bar is shown in the upper-left hand corner to compare with the strength of the identified features.  Five dotted, vertical lines mark some recurrent high-velocity features (at $-2300, -200, 1100, 1700, 3200$ km s$^{-1}$).}	
\end{figure}

\begin{figure}
	\figurenum{16}
	\centering
		\includegraphics[width=0.42\textwidth]{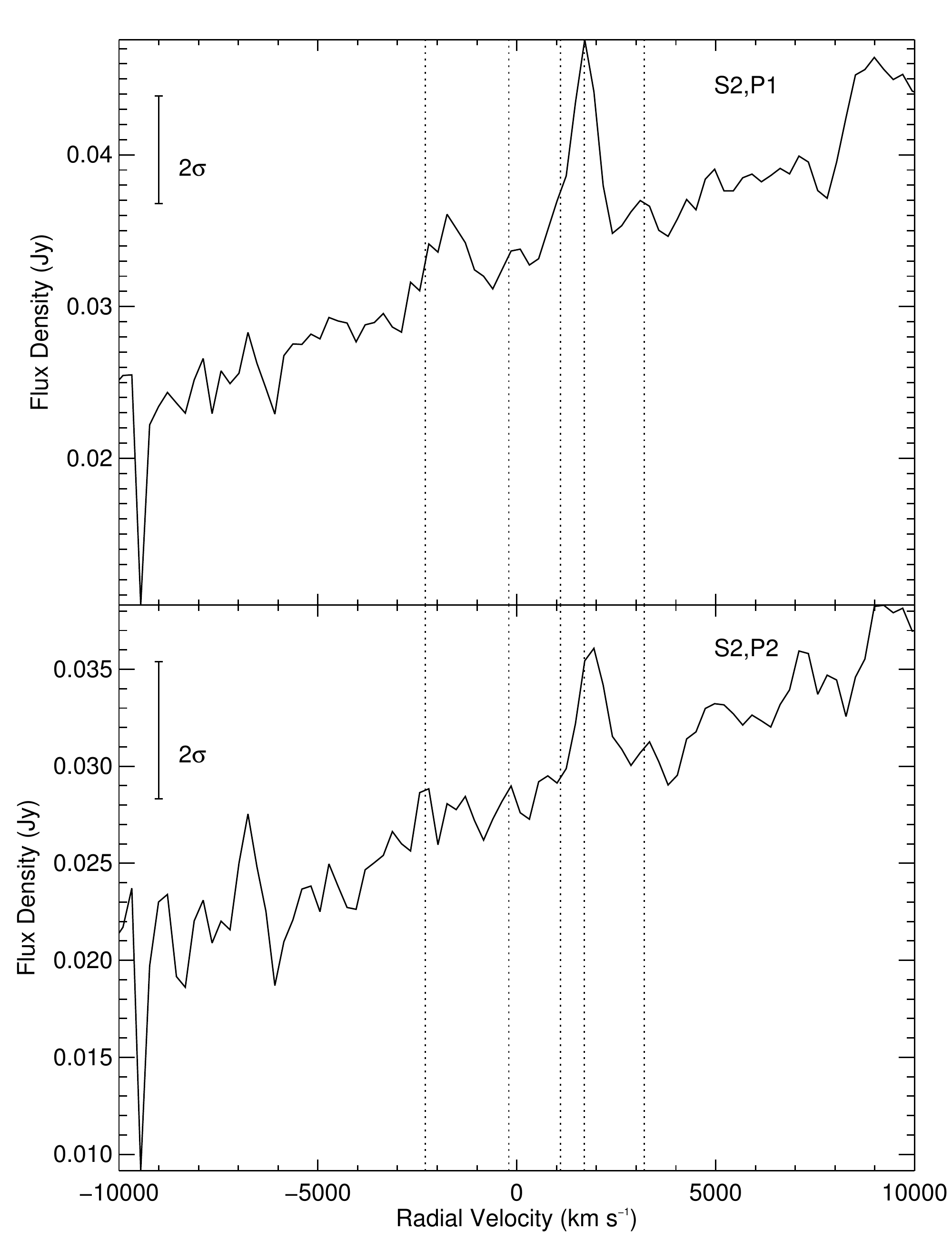}
	\caption{\scriptsize Identified [Ne III] (15.56 $\mu$m) lines in scan 2 (S2) at the southernmost pointings (P1-2).  These lines have been plotted versus relative velocity to demonstrate the presence of high-velocity components in the spectra.  A sample error bar is shown in the upper-left hand corner to compare with the strength of the identified features.  Five dotted, vertical lines mark some recurrent high-velocity features (at $-2300, -200, 1100, 1700, 3200$ km s$^{-1}$).}
\end{figure}

\subsection{Cloudy Models of Purely Photoionized Gas}

We used the spectral synthesis code Cloudy version 13.03 \citep{fer98} to approximate the contribution of radiative excitation to line intensities in the photonionization zone around Cas A.  In models of photoionized ejecta in SNRs, the intensity of the ionizing field is determined by $R_{max}$, the lateral extent of the shock front compared to its thickness \citep{rho09b, raymond79}.  In our models, we set the lateral extent of the shock to $2.4$ pc, in agreement with ejecta models in \citet{lh03} and found the best fit to our mid-infrared spectra with $R_{\max} = 0.81$.  The ionizing radiation field was set by an approximation of X-ray spectra toward Cas A presented in \citet{hl12}.  X-ray spectra toward Cas A peak sharply around $1.9$ keV, and most ejecta knots can be fit with an electron temperature around this value.  Therefore, taking the total X-ray luminosity of Cas A to be $3 \times 10^{36}$ ergs s$^{-1}$ \citep{stewart83}, we approximated the ionizing radiation field in our Cloudy model as a blackbody with a temperature $k_{B} T = 1.9$ keV.  Finally, the abundances were fixed to those calculated for the entire remnant in \citep{hl12}.

Using these approximations, we determined the ejecta column densities that best fit the line intensities in our mid-infrared spectra (Figure 17).  The continuum in this spectrum is dominated by freshly formed dust composed of SiO$_{2}$, Mg protosilicates, and FeO grains \citep{rho08}.  In order to approximate the dust continuum in the region of our fine structure lines, we fit two power-laws independently to the emission shortward and longward of the $21$ micron peak.

\begin{figure}
	\figurenum{17}
	\centering
		\includegraphics[angle=90,width=0.5\textwidth]{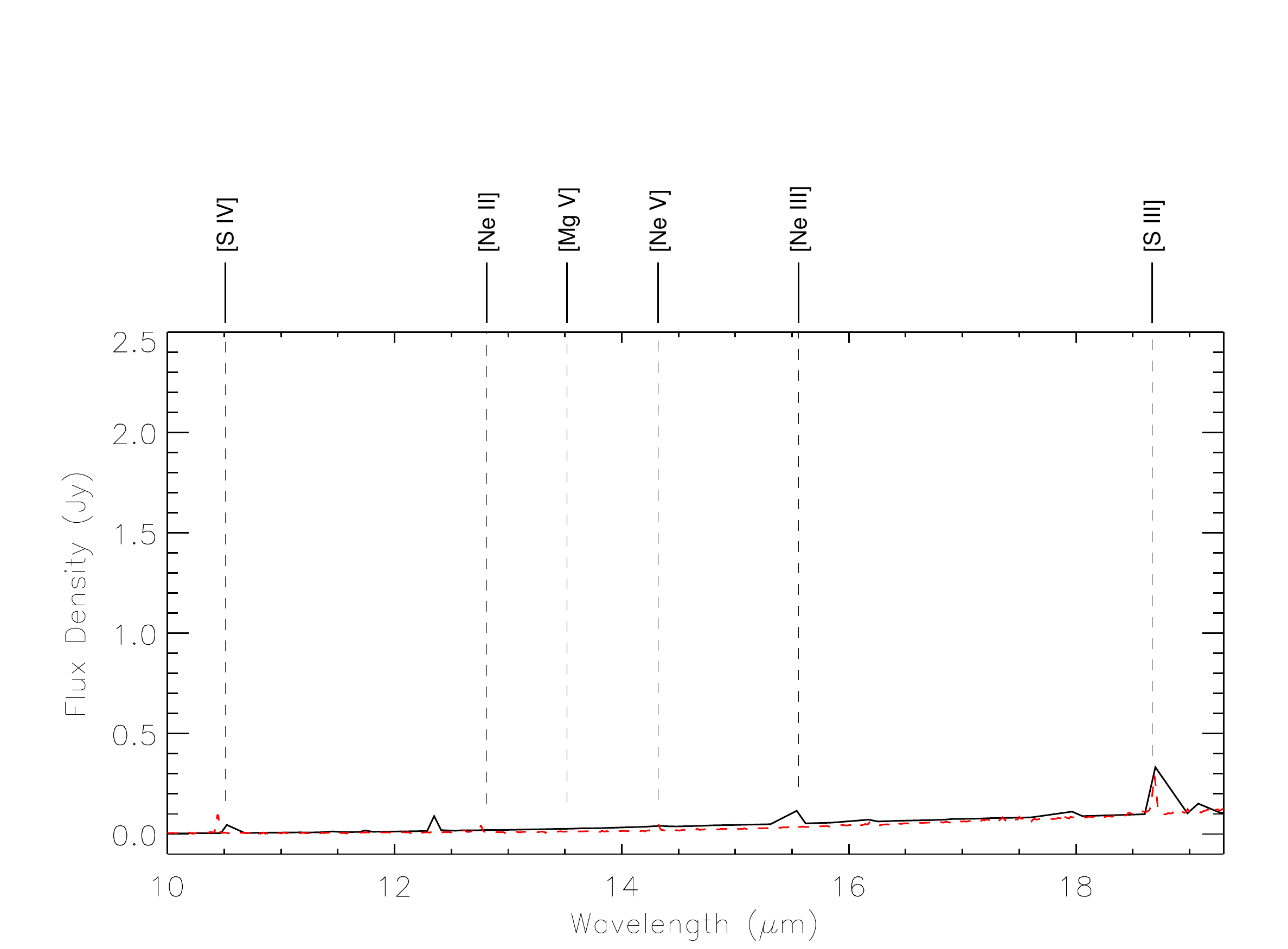}
		\includegraphics[angle=90,width=0.5\textwidth]{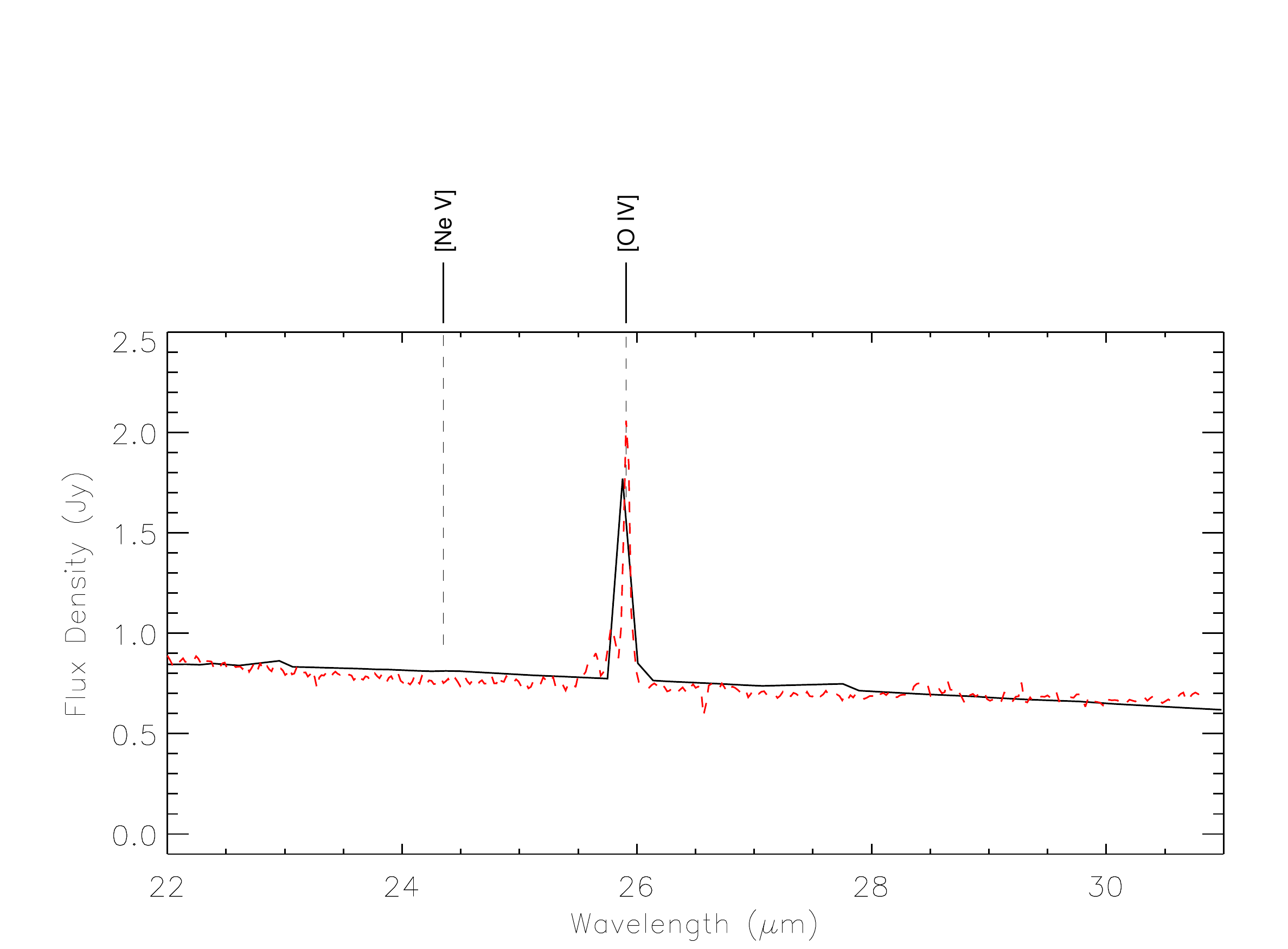}
	\caption{\scriptsize Synthesized spectrum (black, solid line) of gas illuminated by a blackbody to the approximate Cas A gas temperature in X-ray wavelengths fit to our observed spectrum at S2, P1 (red, dashed line).  The line intensities were determined using the CLOUDY photoionization code and the dust continuum emission was fit using two power-laws shortward and longward of the $21$ micron dust continuum peak.}
\end{figure}

We find that the column density of ejecta toward our spectra is approximately $N = 1.1 \times 10^{20}$ cm$^{-2}$ toward the southernmost spectrum (P1) in S2.  Given the IRS slit size and assuming a $3.4$ kpc distance to Cas A, we approximate the total mass of ejecta in our spectra to be $0.05$ M$_{\odot}$.

\section{Alternative Methods for CO Line Broadening}

\subsection{Broadening by Radiative Acceleration}

We can test whether the gas was accelerated by the burst of radiation that was emitted from Cas A when the type IIb event occurred.  From infrared light echoes, \citet{da08} determined that this burst had a luminosity of approximately $L_{r} = 1.5 \times 10^{11}$ L$_{\odot}$ with the maximum intensity at a wavelength around $100$ nm, and lasted for approximately $t_{r} = 40$ days.  Using this information, an assumed distance from the supernova of $d=3$ pc, and an estimate of the cross-sectional cloud area, we can calculate the amount of radiative momentum that could be transferred to a molecular cloud surrounding Cas A and thus estimate the line width of CO produced by the affected gas.

For the size of each molecular cloud, we estimate a surface area oriented toward the source of radiation by assuming that the molecular cloud is approximately a sphere with mass $M$, radius $R$, and column density $\bar{\sigma}$ averaged across the region where we evaluated the cloud mass (Section 3.3).  Thus, $R = \sqrt{M/(\pi \bar{\sigma})}$.  By this method, we obtain radii $1.6, 1.6$ and $1.2$ pc for the western, southwestern, and southeastern clouds, respectively.  The amount of momentum transferred to each cloud by the incident radiation is approximately

\begin{equation}
p_{r} = \int_0^\infty \! \frac{L_{r} t_{r}}{4 \pi d^{2} h \nu} \epsilon \pi R^{2} \frac{2 h}{c} \, \mathrm{d}\nu \approx \frac{\epsilon L_{r} t_{r} R^{2}}{2 d^{2} c}
\end{equation}

\noindent Here we assume the profile of the spectral energy distribution is sharply peaked at 100 nm \citep{da08} such that the majority of the radiation from the Cas A event comes out at this wavelength.  The factor $\epsilon$ is an efficiency with which momentum is transferred to the cloud.  For the limiting case in this paper, we assume that $\epsilon = 1$, although typical values for momentum transfer by scattering of UV photons off of interstellar dust grains suggest that $\epsilon \sim 0.3$ \citep{draine03b}.  The momenta we obtain by substituting the radius of each cloud into this formula are $47, 47, 27 \mbox{ M}_{\odot}$ km s$^{-1}$.  These momenta are two orders of magnitude smaller than required to account for the widths we observe in our molecular clouds (Section 3.3).  It is not possible for radiation from the supernova flash to provide enough acceleration to explain the line widths in our clouds.

\subsection{Broadening by Gravitational Acceleration}

Acceleration in molecular clouds may be a sign of gravitational collapse.  To evaluate the likelihood of this scenario, we have estimated the virial mass of each cloud as

\begin{equation}
M_{\mathrm{vir}} = \frac{5 R \sigma_{w}^{2}}{8 \mathrm{ln(2)} \mathrm{G}}.
\end{equation}

\noindent Here we use $6$ km s$^{-1}$ as the velocity width and the radius is calculated from surface density estimates.  We have assumed that the clouds are spherical and measured their radii along their shortest axis of integrated $^{12}$CO intensity.  The virial masses are then $12000, 12000$ and $9000 M_{\odot}$.  These values are at least an order of magnitude larger than the estimated cloud masses in Section 3.3, thus the clouds are stable against collapse and gravitational acceleration has most likely had no role in producing the observed line widths.


\bibliographystyle{apj}

\end{document}